%% file: manuscript.tex
\documentclass[preprint,12pt,sort&compress,numbers]{elsarticle}

\usepackage{amssymb}
\usepackage{amsmath}
\input{packages}

\input{notations}
\refcolorfalse

\usepackage{lineno}

\journal{Journal of Computational Physics}

\begin{document}

\begin{frontmatter}



\title{Latent Space Dynamics Identification for Interface Tracking with Application to Shock-Induced Pore Collapse}


\author[casc]{Seung Whan Chung\corref{cor1}}
\cortext[cor1]{corresponding author}
\ead{chung28@llnl.gov}
\author[msd]{Christopher Miller}
\author[casc]{Youngsoo Choi}
\author[casc]{Paul Tranquilli}
\author[msd]{H. Keo Springer}
\author[msd]{Kyle Sullivan}

\affiliation[casc]{organization={Center for Applied Scientific Computing, Lawrence Livermore National Laboratory},
  city={Livermore},
  postcode={94550}, 
  state={CA},
  country={US}}
\affiliation[msd]{organization={Material Science Division, Lawrence Livermore National Laboratory},
  city={Livermore},
  postcode={94550}, 
  state={CA},
  country={US}}

\begin{abstract}
\input{sections_abstract_jcp}
\end{abstract}


\begin{highlights}
\item Developed LaSDI-IT, a data-driven latent dynamics modeling framework for physical systems with sharp, moving interfaces.
\item Introduced an interface-aware autoencoder that reconstructs both physical fields and material boundaries.
\item Achieved $<9\%$ prediction error and $10^6\times$ speedup over high-fidelity simulations for shock-induced pore collapse.
\item Demonstrated accurate recovery of key metrics: pore area, hot spot size, and maximum temperature.
\item Reduced required training data size by half using Gaussian process-based greedy sampling while maintaining accuracy.
\end{highlights}


\end{frontmatter}


\input{sections_intro_jcp}
\input{sections_porecollapse}

\input{sections_lasdi_jcp}
\input{sections_results_jcp}
\input{sections_conclusion}

\section*{Acknowledgments}
This work was performed under the auspices of the U.S. Department of Energy
by Lawrence Livermore National Laboratory under contract DE-AC52-07NA27344
and was supported by Laboratory Directed Research and Development funding under project 24-SI-004.
LLNL-JRNL-2006536.






\bibliographystyle{elsarticle-num} 
\bibliography{reference}

\end{document}

%% file: notations.tex
\newif\ifrefcolor
\refcolorfalse

\def\refB#1{\ifrefcolor\textcolor{black!30!green}{#1}\else#1\fi}
\def\refC#1{\ifrefcolor{\color{purple}#1}\else#1\fi}

\newcommand{\YC}[1]{\textcolor{blue}{}}
\newcommand{\KC}[1]{\textcolor{purple}{}}

\newcommand{\kc}[1]{{#1}}

\newcommand{\kctwo}[1]{{#1}}

\newcommand{\En}{\mathrm{En}}
\newcommand{\De}{\mathrm{De}}
\newcommand{\GP}{\mathrm{GP}}

\newcommand{\THE}{T}

\newcommand{\cO}{\mathcal{O}}

\newcommand{\cJ}{\mathcal{J}}
\newcommand{\cD}{\mathcal{D}}

\newcommand{\cN}{\mathcal{N}}
\newcommand{\cT}{\mathcal{T}}

\newcommand{\rR}{\mathbb{R}}

\newcommand{\rT}{\mathbb{T}}
\newcommand{\rP}{\mathbb{P}}

\newcommand{\rD}{\mathbb{D}}

\newcommand{\tT}{\tilde{T}}
\newcommand{\std}{\mathrm{std}}

\newcommand{\bD}{\mathbf{D}}

\newcommand{\bT}{\mathbf{T}}
\newcommand{\bW}{\mathbf{W}}
\newcommand{\bZ}{\mathbf{Z}}
\newcommand{\bb}{\mathbf{b}}

\newcommand{\bq}{\mathbf{q}}
\newcommand{\bx}{\mathbf{x}}

\newcommand{\bz}{\mathbf{z}}

\newcommand{\bp}{\mathbf{p}}

\newcommand{\bPhi}{\bm{\Phi}}
\newcommand{\bXi}{\bm{\Xi}}
\newcommand{\bphi}{\bm{\phi}}

\def\deg#1{#1^{\circ}}
\def\En{\mathrm{En}}
\def\De{\mathrm{De}}

\pgfplotsset{select coords between index/.style 2 args={
    x filter/.code={
        \ifnum\coordindex<#1\fi
        \ifnum\coordindex>#2\fi
    }
}}

%% file: sections_abstract_jcp.tex
Capturing sharp, evolving interfaces remains a central challenge in
  reduced-order modeling, especially when data is limited and the system
  exhibits localized nonlinearities or discontinuities.
  We propose LaSDI-IT
  (Latent Space Dynamics Identification for Interface Tracking), a data-driven
  framework that combines low-dimensional latent dynamics learning with
  explicit interface-aware encoding to enable accurate and efficient modeling
  of physical systems involving moving material boundaries. At the core of
  LaSDI-IT is a revised autoencoder architecture that jointly reconstructs the
  physical field and an indicator function representing material regions or
  phases, allowing the model to track complex interface evolution without
  requiring detailed physical models or mesh adaptation. The latent dynamics
  are learned through linear regression in the encoded space and generalized
  across parameter regimes using Gaussian process interpolation with greedy
  sampling. We demonstrate LaSDI-IT on the problem of shock-induced pore
  collapse in high explosives, a process characterized by sharp temperature
  gradients and dynamically deforming pore geometries. The method achieves
  relative prediction errors below 9\% across the parameter space, accurately
  recovers key quantities of interest such as pore area and hot spot formation,
  and matches the performance of dense training with only half the data.
  \kc{This latent dynamics prediction was $10^6$ times faster than
  the conventional high-fidelity simulation,
  proving its utility for multi-query applications.}
  These results highlight LaSDI-IT as a general, data-efficient framework for
  modeling discontinuity-rich systems in computational physics, with potential
  applications in multiphase flows, fracture mechanics, and phase change
  problems.

%% file: sections_intro_jcp.tex
\section{Introduction}\label{sec:intro}

Many physical systems of scientific and engineering interest are governed by
complex dynamics involving sharp, moving interfaces. These interfaces arise in
a wide range of applications, including shock propagation in compressible
flows, phase transitions in heat and mass transfer, multiphase and reactive
flows, fracture and crack propagation in solids, and front evolution in porous
media.
\par
\kctwo{
Shock-induced pore collapse in high explosives (HEs)
is one of the examples that requires accurate resolution of sharp interface evolution.
Pores in HE materials are defects in the energetic crystal and crystal-crystal interface which provide little resistance to the
oncoming shock wave.
These pores are common locations for localized temperature spikes---known as hot spots---to form~\cite{menikoff2004pore, springer2017effects},
which may transition into a detonation under appropriate
conditions~\cite{bowden1985initiation,fried2012role, tarver1996critical,field1992hot}.
Pore collapse is thus one of the key mechanisms
that trigger the ignition of HEs under shock loading.
The resulting plastic deformation and jetting of the HE
undergoing pore collapse has been widely studied over the years to attempt to
predict hot spot formation~\cite{austin2014modeling, miller2019ignition,
miller2021probabilistic, kapahi2015three}.
}
\par
\kctwo{
Accurate modeling of such systems often requires high-fidelity numerical
solvers that finely resolve interface geometry and complex nonlinear evolution~\cite{noble2017ale3d}.
This in turn results in substantial computational cost,
which makes high-fidelity solvers impractical for large-scale simulations,
real-time prediction, design optimization, and uncertainty quantification.
}
\par
Reduced-order modeling (ROM) offers a promising strategy for alleviating the
computational burden by constructing low-dimensional surrogate models that
approximate key features of the underlying physical system. Traditional ROM
techniques, such as projection-based methods~\cite{berkooz1993proper,
rozza2008reduced, safonov2002schur, lee2020model, kim2022fast, diaz2024fast,
zanardi2024scalable, amsallem2015design, choi2020gradient,
carlberg2018conservative, choi2019space, choi2021space, hoang2021domain,
mcbane2021component, copeland2022reduced, kim2020efficient, choi2020sns,
choi2019accelerating, kim2021efficient, cheung2023local, lauzon2024s,
mcbane2022stress, chung2024train, tsai5134633local}, have been widely used to
accelerate simulations by
projecting governing equations onto a reduced basis identified from data or
physical insight. \kctwo{However, these methods are typically intrusive, requiring
in-depth manipulation of the physics equation and its discretized solver.
On the other hand, non-intrusive ROM approaches recently have gained
traction due to their flexibility and solver-agnostic nature.} These methods
learn low-dimensional embeddings and surrogate dynamics directly from
simulation or experimental data, using tools such as dynamic mode
decomposition (DMD)~\cite{schmid2010dynamic, rowley2009spectral, tu2013dynamic,
proctor2016dynamic}, autoencoders~\cite{kadeethum2022non, tran2024weak,
park2024tlasdi}, sparse regression~\cite{brunton2016discovering}, operator
inference~\cite{peherstorfer2016data, mcquarrie2021data,
mcquarrie2023nonintrusive}, and neural operators~\cite{li2020fourier,
kovachki2023neural, lu2019deeponet}.
\par
Despite their potential, these ROM approaches
face fundamental challenges when applied to systems with sharp interfaces.
Standard reduced bases \kctwo{of projection-based methods}—linear or nonlinear—often struggle
to capture discontinuities or steep gradients that characterize
interface-dominated phenomena.
\kctwo{Non-intrusive approaches likewise suffer from capturing discontinuities.}
Neural networks, in particular, are known to exhibit \emph{spectral
bias}~\cite{rahaman2019spectral}, meaning they preferentially learn smooth,
low-frequency components of the data. As a result, high-frequency features,
such as shocks, phase boundaries, or fracture tips, are smoothed out or poorly
reconstructed, leading to degraded model accuracy and generalization.
This issue becomes even more pronounced in data-scarce regimes, which are
common in many scientific domains where high-fidelity simulations or physical
experiments are expensive. In such settings, the inability to capture and
evolve sharp interface features can severely limit the utility of ROMs for
downstream tasks such as control, optimization, and forecasting.
\par
\kctwo{
Several data-driven surrogate modeling methods have been
recently proposed for the shock dynamics of heterogeneous
materials~\cite{nguyen2023physics, li2023mapping, springer2023simulating,
cheung2024data}.  However, due to limited training data, the accuracy of
instantaneous temperature evoluation is often
compromised~\cite{nguyen2023physics}, or predictions are limited to
post-shock temperature field~\cite{li2023mapping} or reaction
rate~\cite{springer2023simulating}.  Recent work employing DMD for
shock-induced pore collapse successfully predicts instantaneous evolution of
temperature fields~\cite{cheung2024data}.  However, the linear nature of DMD
necessitates training multiple DMD models for consecutive short time windows,
which reduces the generalizability of the trained model.
}
\par
To address this challenge, we introduce a data-driven surrogate modeling
framework tailored for systems with sharp, moving interfaces: \emph{Latent
Space Dynamics Identification for Interface Tracking (LaSDI-IT)}. LaSDI-IT
extends the recently developed LaSDI framework~\cite{fries2022lasdi, he2023glasdi,
bonneville2024gplasdi}, which combines autoencoder-based dimensionality
reduction with explicit modeling of latent dynamics via ordinary differential
equations (ODEs). The key innovation in LaSDI-IT is a revised autoencoder
architecture that jointly reconstructs both the physical state and a binary
indicator function that tracks the spatial extent of the active (non-interface)
domain. This design enables the latent representation to faithfully encode
interface dynamics and maintain reconstruction accuracy \kctwo{near the interface}.
To ensure parametric generalization, LaSDI-IT models the latent dynamics as
parametric ODEs and uses Gaussian process regression with greedy sampling to
interpolate across parameter space while minimizing the number of required
training simulations. This makes the method well-suited for multi-query
scenarios in data-limited regimes.
\par
As a representative test case, we apply LaSDI-IT to the shock-induced pore
collapse problem.
However, the scope of LaSDI-IT
is much broader—it is applicable to any problem where interface fidelity is
essential for reduced modeling.
\par
The remainder of this paper is organized as follows.
\kctwo{
Section~\ref{sec:porecollapse}
describes high-fidelity simulation setup and training data generation
for the shock-induced pore collapse as an example physics problem with a moving interface.
Section~\ref{sec:lasdi} introduces the proposed LaSDI-IT framework,
with Section~\ref{sec:ae} highlighting the modifications necessary for robust
interface tracking.
}
Section~\ref{sec:results} presents numerical results demonstrating
the improved reconstruction accuracy and predictive performance of LaSDI-IT
over standard approaches. Finally, Section~\ref{sec:conclusion} concludes with
a discussion on the generalizability of the framework, current limitations, and
directions for future research.

%% file: sections_porecollapse.tex
\section{Shock-induced pore collapse}\label{sec:porecollapse}
\input{sections_ale3d}

%% file: sections_ale3d.tex
Pore collapse simulations are performed using the Arbitrary Lagrangian–Eulerian
multiphysics code ALE3D~\cite{noble2017ale3d}. The equation of state (EOS) for
the unreacted high explosive (HE) is computed using the thermochemical
equilibrium code Cheetah~\cite{fried2002exp6}, and the material strength is
represented using the Johnson–Cook constitutive
model~\cite{johnson1983constitutive}. Further details on the EOS formulation,
strength modeling, and their application to HE materials can be found in
previous studies~\cite{springer2017effects, springer2023simulating,
cheung2024data}.
\par
These simulations do not incorporate chemical kinetics; instead, the focus is
on capturing temperature increases in the HE resulting from mechanical energy
dissipation. This choice allows us to isolate the key physical mechanisms
responsible for hot spot formation, which are critical for training the latent
dynamics model. Moreover, since temperature rise is the dominant driver of
chemical decomposition and ignition, accurately predicting localized heating
and hot spot development provides strong indicators of whether ignition is
likely to occur.
\par
To this end, we define a temperature threshold of $800\,\mathrm{K}$, above
which any zonal element is considered part of a developing hot spot. This
threshold is selected to lie well above the typical bulk heating temperature
induced by shock loading at similar velocities (approximately $600\,\mathrm{K}$), yet below the
localized temperature spikes that occur near the pore during collapse (often
exceeding $1000\,\mathrm{K}$).
\par
In this study, pore collapse dynamics are modeled with respect to geometric variations in pore shape.
This focus on geometry is chosen for demonstration purposes,
although other physical parameters—such as shock speed or material composition—could also be considered.
The pore is idealized as an oval with a fixed minor axis of $1\,\mu\mathrm{m}$,
while the major axis varies between $1\,\mu\mathrm{m}$ and $1.2\,\mu\mathrm{m}$
and is oriented at an angle between $0^\circ$ and $20^\circ$ from the direction of the shock.
These two parameters—the orientation angle $p_1$ and the major axis length $p_2$—define the physical parameter vector $\bp = (p_1, p_2)$, as illustrated in Figure~\ref{fig:pore-params}.
\begin{figure}[tbph]
    \input{figures_pore_params.tex}
    \caption{Geometric parameterization $\bp = (p_1, p_2)$ of the elliptical pore.}
    \label{fig:pore-params}
\end{figure}
The simulations involve a $3\,\mu\mathrm{m}$ thick aluminum flyer traveling at $2km/s$
impacting a slab of HE containing a single pore, generating a shock pressure of
approximately $11.5\,\mathrm{GPa}$. The simulation runs for $3\,\mathrm{ns}$,
sufficient for the shock to traverse the pore and for the temperature field to
reach a quasi-steady state. A total of 300 time steps are computed using a time
step size of $\Delta t = 0.01\,\mathrm{ns}$. The computational domain is a
$15\,\mu\mathrm{m} \times 15\,\mu\mathrm{m}$ square with uniform mesh spacing
of $30\,\mathrm{nm}$ to ensure accurate resolution of far-field boundary
conditions.
\kc{
A simulation for each parameter point takes about 10 minutes
\refB{on 112 Intel Sapphire Rapids CPU cores
at $2\,\mathrm{GHz}$ clock speed with $2\,\mathrm{GB}$ memory per core}}~\cite{dane}.
\par
For training the surrogate model, we extract a focused subdomain $\rD = [0,
5\,\mu\mathrm{m}] \times [0, 2\,\mu\mathrm{m}]$ that captures the critical
evolution of the HE temperature field near the pore. Additionally, data from
the first $1.15\,\mathrm{ns}$ are excluded, as this interval consists primarily
of shock advection before interaction with the pore. The subdomain $\rD$
contains $N_x = 66 \times 167 = 11{,}022$ spatial grid points. 
For each parameter setting $\bp$, the corresponding time series of the HE temperature field is denoted by $\bT(\bp)$:
\begin{equation}\label{eq:T}
    \bT(\bp)
    =
    \begin{pmatrix}
    T_0(\bp)
    &
    T_1(\bp)
    &
    \cdots
    &
    T_{N_t}(\bp)
    \end{pmatrix}
    \in
    \rR^{N_x\times (N_t+1)}.
\end{equation}
The full training dataset, consisting of simulations at parameter points $\rP = \{\bp_1, \bp_2, \ldots, \bp_{N_p}\}$, is represented as:
\begin{equation}\label{eq:full-training-T}
\rT = \{ \bT(\bp_1), \bT(\bp_2), \ldots, \bT(\bp_{N_p}) \}.
\end{equation}
\par
\begin{figure}[tbph]
    \input{figures_HE_temp.tex}
    \caption{(a) Temperature field $\THE$ of high explosive (HE) material with $0K$ indicating non-HE region, and (b) $\THE$ with mask $\THE>0$.}
    \label{fig:he-temp}
\end{figure}
Figure~\ref{fig:he-temp} presents a snapshot of the HE temperature field at $t
= 0.35\,\mathrm{ns}$ for a pore configuration with $\bp = (1.2\,\mu\mathrm{m},
20^\circ)$. The pore region is represented by zero temperature
($0\,\mathrm{K}$), marking a discontinuous interface with the surrounding HE
material.
\kctwo{
To characterize hot spot formation and assess detonation sensitivity,
we focus on three key physical metrics: (1) the hot spot area
with the threshold temperature $800K$,
\begin{equation}\label{eq:A-hotspot}
A_{hotspot}(t_k; \bp) =
\int_{\rD} \mathbf{1}_{T_k(\bx;\bp)>800}(\bx)\;d\bx^2,
\end{equation}
(2) the pore area,
\begin{equation}\label{eq:A-pore}
A_{pore}(t_k; \bp) =
\int_{\rD} \mathbf{1}_{T_k(\bx;\bp)=0}(\bx)\;d\bx^2,
\end{equation}
and (3) the maximum temperature over time,
\begin{equation}\label{eq:max-temp}
T_{k, max}(\bp) = \max_{\bx}T_k(\bx; \bp).
\end{equation}
}

%% file: figures_pore_params.tex
\begin{tikzpicture}[
    font=\small,
    spy using outlines={circle,black,magnification=6,size=1.5cm, connect spies}
    ]
    
    \pgfplotsset{set layers=standard}
    \begin{groupplot}[
        group style={
            group name = my plots,
            group size= 1 by 1,
            xlabels at =edge bottom,
            horizontal sep=3.5cm,
            vertical sep=2.cm,
        },
        name=chung,
        height = 0.27\textwidth,
        width = 0.55\textwidth,
    ]    

        \nextgroupplot[
            enlarge x limits={false, abs value = 0mm},
            ylabel={$x_2$ ($\mu m$)},
            xlabel={$x_1$ ($\mu m$)},
            tick scale binop ={\times},
            xmin=0,xmax=5,ymin=0,ymax=2,
            point meta min=323., point meta max=606,
            colorbar, colormap/viridis,
            colorbar style={
                font=\scriptsize,
                xticklabel pos=upper,
                scaled y ticks=false,
                /pgf/number format/precision=4,
                xlabel={$T$ ($K$)},
            },
        ]
        
            \edef\imagepath{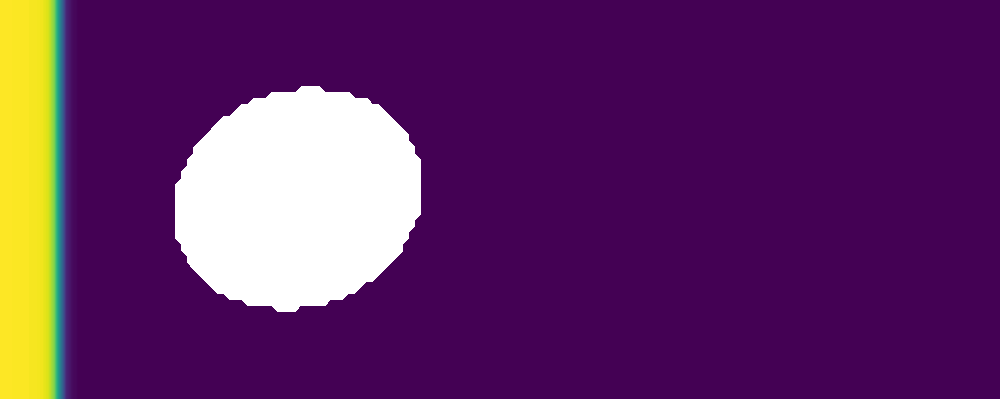}
            \addplot graphics[xmin=0,xmax=5,ymin=0,ymax=2]{\imagepath};

            \draw[solid, red, line width=1.0, <->,] (axis cs: 2.110, 1.22) -- (axis cs: 0.889, 0.777) node[anchor=south west, yshift=4pt] {$p_2$};

            \draw[solid, red, line width=0.5, -,] (axis cs: 1.7, 1) -- (axis cs: 3, 1);
            \draw[solid, red, line width=0.5, -,] (axis cs: 2.157, 1.239) -- (axis cs: 2.909, 1.513);

            \draw[solid, red, line width=1.0, <->] (axis cs: 2.8, 1) arc (0:18:1.5cm) node[anchor=north west, xshift=5pt,] {$p_1$};
    
    \end{groupplot}
    \node[above = 0.1cm of my plots c1r1.north,
        anchor=south, xshift=-.5cm,
    ] {$\Longrightarrow$ shock traveling direction};
\end{tikzpicture}
    

%% file: figures_HE_temp.tex
\begin{tikzpicture}[
    font=\small,
    spy using outlines={circle,black,magnification=6,size=1.5cm, connect spies}
    ]
    
    \pgfplotsset{set layers=standard}
    \begin{groupplot}[
        group style={
            group name = my plots,
            group size= 2 by 1,
            xlabels at =edge bottom,
            horizontal sep=3.5cm,
            vertical sep=2.cm,
        },
        name=chung,
    ]    

        \nextgroupplot[
            height = 0.22\textwidth,
            width = 0.45\textwidth,
            enlarge x limits={false, abs value = 0mm},
            ylabel={$x_2$ ($\mu m$)},
            xlabel={$x_1$ ($\mu m$)},
            tick scale binop ={\times},
            xmin=0,xmax=5,ymin=0,ymax=2,
            point meta min=0., point meta max=790,
            colorbar, colormap/viridis,
            colorbar style={
                font=\scriptsize,
                xticklabel pos=upper,
                scaled y ticks=false,
                /pgf/number format/precision=4,
                xlabel={$T$ ($K$)},
            }
        ]
        
            \edef\imagepath{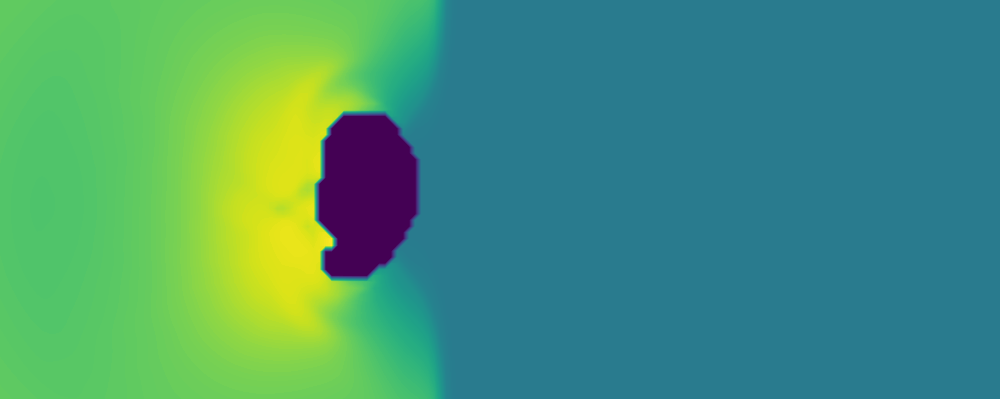}
            \addplot graphics[xmin=0,xmax=5,ymin=0,ymax=2]{\imagepath};


        \nextgroupplot[
            height = 0.22\textwidth,
            width = 0.45\textwidth,
            enlarge x limits={false, abs value = 0mm},
            ylabel={$x_2$ ($\mu m$)},
            xlabel={$x_1$ ($\mu m$)},
            tick scale binop ={\times},
            xmin=0,xmax=5,ymin=0,ymax=2,
            point meta min=323., point meta max=790,
            colorbar, colormap/viridis,
            colorbar style={
                font=\scriptsize,
                xticklabel pos=upper,
                scaled y ticks=false,
                ytick={323., 400, 600, 790},
                yticklabels={323, 400, 600, 790},
                /pgf/number format/precision=4,
                xlabel={$T$ ($K$)},
            }
        ]
        
            \edef\imagepath{data_Xtrain36.pure_zone.case35.timestep35.png}
            \addplot graphics[xmin=0,xmax=5,ymin=0,ymax=2]{\imagepath};
    
    \end{groupplot}
    \node[below = 1.5cm of my plots c1r1.south west,
        anchor=west,
    ] {(a) Raw temperature field};
    \node[below = 1.5cm of my plots c2r1.south west,
        anchor=west,
    ] {(b) $T>0$};
\end{tikzpicture}
    

%% file: sections_lasdi_jcp.tex
\section{Methodology}\label{sec:lasdi}

\subsection{Overall framework}
Given a time series of a state field at a specific parameter point $\bp \in
\mathbb{R}^{N_{\mu}}$, \kctwo{for example, represented as $\bT(\bp)$ in (\ref{eq:T})},
the objective of the Latent Space Dynamics Identification (LaSDI) framework is
to identify a low-dimensional latent variable $\bz \in \mathbb{R}^{N_z}$, with
$0 < N_z \ll N_x$, and its corresponding dynamics, expressed as $\dot{\bz} =
\mathcal{L}[\bz; \bp]$, that can effectively approximate the evolution of the
high-dimensional state field $\bT(\bp)$.
\par
\begin{figure*}[tb]
    \input{figures_schematic.tex}
    \caption{Schematic diagram of latent space dynamics identification framework.}
    \label{fig:schematic}
\end{figure*}
Several variants of LaSDI have been proposed to accommodate different modeling
goals and data regimes. In this study, we adopt the GPLaSDI
framework~\cite{bonneville2024gplasdi}, a non-intrusive, data-driven
formulation of LaSDI that combines Gaussian process interpolation with greedy
sampling to enable efficient surrogate modeling across parameterized systems.
\kctwo{
Figure~\ref{fig:schematic} illustrates the overall GPLaSDI framework.
First, an autoencoder (AE) is employed to encapsulate essential features of a state field $T(\bp)$
onto a low-dimensional nonlinear manifold.
The evolution of the latent space variable $\bz$, as captured by AE,
is represented using sparse identification of nonlinear dynamics (SINDy)~\cite{brunton2016discovering,fries2022lasdi}.
We further assume parametric dependence of the latent space dynamics,
and predict the dynamics at \refC{an} unseen parameter point $\bp$
based on the training parameter points $\{\bp_1, \bp_2, \ldots, \bp_{N_p}\}$,
using Gaussian process.
Each component of the proposed framework will be introduced subsequently.
}
For a detailed overview of the original LaSDI methodology and its extensions,
we refer the reader to previous
works~\cite{fries2022lasdi,bonneville2024gplasdi}.

\subsection{\kctwo{Standard} autoencoder}\label{sec:standard-ae}

Typically, an AE consists of two neural-networks (NNs): an encoder and a decoder,
\begin{subequations}
    \begin{equation}\label{eq:en}
    \bz = \En(T;\theta_{en})
    \end{equation}
    \begin{equation}\label{eq:de}
    T = \De(\bz;\theta_{de}),
    \end{equation}
\end{subequations}
with their weight and biases denoted as $\theta_{en}$ and $\theta_{de}$.
The encoder compresses the high-dimensional state field into a latent space variable $\bz$,
while the decoder reconstructs the state field from the latent space variable.
\par
A multilayer perceptron (MLP) with depth $L$ is used for both
the encoder and decoder, defined as
\begin{equation}\label{eq:mlp}
NN(\bq;\theta) = \left( \cT_{L+1}\circ \sigma \circ \cT_L \circ \cdots \circ \sigma \circ \cT_1 \right)(\bq;\theta),
\end{equation}
for a general vector $\bq\in\rR^{d}$,
where each layer $\cT_k: \rR^{d_{k-1}}\to\rR^{d_k}$ is an affine function
\begin{equation}
\bq_k \equiv \cT_k(\bq_{k-1}) = \bW_k\bq_{k-1} + \bb_k,
\end{equation}
and $\sigma$ is an activation function.
The NN parameter is composed of weight matrices and bias vectors of all layers,
i.e. $\theta = \{\bW_1, \bb_1, \ldots, \bW_{L+1}, \bb_{L+1}\}$.
\par
To effectively train an autoencoder,
scaling of the training data is often required.
This is particularly true for pore collapse problems,
where drastic change in state occurs in a localized region near the pore during the hot spot development.
To ensure uniform variance in state across the grid space,
the state field is scaled with temporal average $\overline{T}$ and standard deviation $\std[T]$,
\begin{equation}\label{eq:T-scale}
\tT = \frac{T - \overline{T}}{\std[T]},
\end{equation}
where $\overline{T}$ and $\std[T]$ are evaluated over the entire training data $\rT$,
\begin{subequations}\label{eq:T-scale0}
\begin{equation}
\overline{T} = \frac{1}{N_pN_t}\sum_{p}^{N_p}\sum_k^{N_t}T_k(\bp_p)
\end{equation}
\begin{equation}
\std[T] =
\left(
    \frac{1}{N_pN_t}\sum_{p}^{N_p}\sum_k^{N_t}T_k^2(\bp_p) - \overline{T}^2
\right)^{1/2}.
\end{equation}    
\end{subequations}
This scaling helps the autoencoder learn the features of state field uniformly across the grid points.
\par
The neural network (NN) parameters of the autoencoder are trained to minimize
the reconstruction loss of the scaled training data:
\begin{equation}\label{eq:J-ae0}
\cJ_{AE} = \frac{1}{N_pN_tN_x}\sum_{p=1}^{N_p}\sum_{k=1}^{N_t}
\left\Vert \tT_k(\bp_p) - \De(\En(\tT_k(\bp_p); \theta_{en}); \theta_{de}) \right\Vert^2,
\end{equation}
where $\En$ and $\De$ denote the encoder and decoder networks, respectively,
and $\theta_{en}, \theta_{de}$ represent their corresponding trainable
parameters. The input fields $\tT_k(\bp_p)$ are temporally scaled using the
average and standard deviation as defined in (\ref{eq:T-scale0}). For
untrained models, the reconstruction loss (\ref{eq:J-ae0}) is typically on the
order of $\cO(1)$.
\par
An effective autoencoder is critical for LaSDI, as it enables accurate
reconstruction of high-dimensional states from a compact latent representation
$\bz$, while also facilitating efficient and interpretable latent dynamics
learning. However, designing an autoencoder for systems with sharp, moving
interfaces requires special consideration. These systems often contain
localized discontinuities or rapidly evolving geometrical structures, which
standard neural networks tend to smooth out due to spectral bias.
\par
\kctwo{
There are many novel NN architectures and training methods proposed to tackle
the spectral bias of NN, such as multi-stage NN~\cite{wang2024multi} and
Fourier feature \refC{mapping}~\cite{tancik2020fourier}.  However, our goal in this
study is not to capture high-frequency modes against the spectral bias of NN,
but to track the location of moving interface.
}
\par
Therefore, the autoencoder architecture, loss function, and preprocessing—such
as scaling and masking—must be adapted to the nature of the data and the
physics of the problem. For problems involving sharp interface dynamics (e.g.,
shocks, fractures, phase boundaries), tailored designs are often necessary to
preserve interface fidelity and minimize reconstruction artifacts.  One such
design is proposed in detail \kctwo{in the subsequent section,} and its underlying
principles are applicable to a broader class of interface-dominated systems.

\input{sections_ae_jcp}

\subsection{Latent dynamics identification}
The evolution of the latent space variable $\bz$, as captured by AE,
is represented as a system of ordinary differential equations (ODEs)
with a right-hand side composed of candidate terms~\cite{brunton2016discovering,fries2022lasdi},
\begin{equation}\label{eq:lasdi}
\dot{\bz} = \bXi(\bp)^{\top}\bPhi(\bz)^{\top},
\end{equation}
with $\bPhi$ a dictionary for candidate terms and $\bXi$ their corresponding coefficients.
For example of a general quadratic ODE system,
\begin{equation}
\bPhi(\bz) =
\begin{pmatrix}
1 & \bz^{\top} & z_1^2 & z_2^2 & \ldots & z_1z_2 & \ldots & z_{N_z-1}z_{N_z}
\end{pmatrix}.
\end{equation}
\refB{
Although SINDy supports general nonlinear formulations,
we restrict the latent dynamics to a linear form for the shock-induced pore collapse examples considered in this study.
This choice is motivated by two factors.
First,
restricting the dynamics to a linear form substantially reduces the number of coefficients to be fitted,
thereby mitigating the risk of overfitting without resorting to sparse regression.
This in turn avoids the potential slow-down in multi-objective training that would be incurred by an additional sparsity regularization term.
Second,
although the temperature field exhibits complex spatial features,
its evolution is predominantly one-directional as the shock propagates through the domain.
Once the shock has passed and the pore has fully collapsed,
the temperature field settles into a quasi-steady state.
This observation suggests that the evolution of the temperature field
can be adequately represented by linear dynamics with a single fixed point,
while the autoencoder captures the nonlinear solution manifold.
Nevertheless, the proposed framework can be extended to nonlinear forms in a straightforward manner.
}
\par
The coefficients of the
linear latent dynamics are estimated from data solving \refC{a} regression problem.
From a time series of state field $\bT(\bp)$, we obtain the time series of
corresponding latent variables encoded by AE,
\begin{equation}
\bZ \equiv \En(\bT) =
\begin{pmatrix}
\bz_1 & \bz_2 & \ldots & \bz_{N_t}
\end{pmatrix}
\in\rR^{N_z\times N_t}.
\end{equation}
The time derivative of this data $\bZ$ is evaluated using finite-difference (FD) approximation,
\begin{equation}
\dot{\bZ}
\approx
\frac{1}{\Delta t}\bZ\bD_t^{\top},
\end{equation}
where $\bD_t\in\rR^{N_t\times N_t}$ is a FD stencil operator using summation-by-part 2nd-4th order stencil~\cite{kreiss1974finite,hicken2013summation}.
Given $\bZ$ and $\dot{\bZ}$,
we seek the coefficient $\bXi\in\rR^{(N_z+1)\times N_z}$ that best-minimizes the residual for the governing ODE system,
\begin{equation}
\bXi = \arg\min\limits_{\tilde{\bXi}} \left\Vert \frac{1}{\Delta t}\bD_t\bZ^{\top} - \bPhi(\bZ)\tilde{\bXi} \right\Vert^2,
\end{equation}
with the candidate right-hand side term for training data
\begin{equation}
\bPhi(\bZ) = 
\begin{pmatrix}
\mathbf{1} & \bZ^{\top}
\end{pmatrix}
\in \rR^{N_t\times (N_z+1)}.
\end{equation}
This minimizer can be obtained using the pseudo-inverse of $\bPhi$,
\begin{equation}\label{eq:xi}
\bXi(\bp) = \frac{1}{\Delta t}\bPhi(\bZ(\bp))^{\dagger}\bD_t\bZ(\bp)^{\top}.
\end{equation}
\par
In the GPLaSDI framework, the autoencoder is trained to minimize the remaining residual of latent dynamics as well,
\begin{equation}\label{eq:J-ld}
\cJ_{LD} = \frac{1}{N_p}\sum_p^{N_p}\left\Vert \frac{1}{\Delta t}\bD_t\bZ(\bp_p)^{\top} - \bPhi(\bZ(\bp_p))\bXi(\bp_p) \right\Vert^2,
\end{equation}
where the minimizer (\ref{eq:xi}) is plugged into the ODE residual.
To avoid excessively large values in the SINDy coefficients, which can result
in ill-conditioned latent ODE systems, an $\ell_2$ regularization term is added
to the training objective. The total loss function thus combines the
reconstruction loss $\cJ_{AE}$, the latent dynamics residual $\cJ_{LD}$, and a
regularization penalty on the coefficient norm:
\begin{equation}\label{eq:J}
  \cJ = \cJ_{AE} + \beta_1 \cJ_{LD} + \beta_2 \| \bXi \|_2^2,
\end{equation}
where $\beta_1$ and $\beta_2$ are hyperparameters that balance the
contributions of the dynamics residual and the regularization term.

Also, \refC{choosing the} optimal value for the relative weight $\beta_1$ is entirely problem-specific and requires hyper-parameter tuning,
though we recommend tuning it in unit of $\Delta t^2$.
The rationale is that, with (\ref{eq:xi}),
the latent dynamics loss (\ref{eq:J-ld}) is always multiplied by a factor of $\Delta t^{-2}$.
Thus, this approach essentially non-dimensionalizes $\cJ_{LD}$ regardless of the unit of $\Delta t$,
reducing the effort in hyper-parameter tuning.

\subsection{Parametric interpolation via Gaussian-process}
For the training parameter points $\rP=\{\bp_1, \bp_2, \ldots, \bp_{N_p}\}$,
the full training dataset is given as
$\rT = \{\bT(\bp_1), \bT(\bp_2), \ldots, \bT(\bp_{N_p})\}$ in (\ref{eq:full-training-T}).
Once the dynamics coefficients $\bXi$ are identified for these training parameter points,
the next step is to estimate the coefficients at \refC{an} unseen parameter point, i.e. $\bXi(\bp^*)$,
based on the `measurements' $\cD=\{\bXi(\bp_1), \bXi(\bp_2), \ldots, \bXi(\bp_{N_p})\}$.
\par
Gaussian process (GP) is employed for this purpose~\cite{rasmussen2003gaussian}.
For an entry of the coefficient matrix, $\Xi_{ij}$,
the measurement data $\cD_{ij}=\{\Xi_{ij}(\bp_1), \ldots, \Xi_{ij}(\bp_{N_p})\}\in\rR^{N_p}$
is given at the training parameter points $\rP$.
Each $\Xi_{ij}$ is treated as an uncertain variable that follows an independent Gaussian process,
\begin{equation}\label{eq:gp}
\Xi_{ij}(\bp)
\sim
\GP\!\left[\overline{\Xi}_{ij}(\bp), K(\bp, \bp')\right]
+
\epsilon.
\end{equation}
The GP assumes a prior mean model $\overline{\Xi}_{ij}(\bp): \rR^{N_p}\to \rR$
and uses a covariance kernel $K(\bp, \bp'): \rR^{N_p}\times\rR^{N_p}\to \rR$
to quantify the relationship between parameter points.
Additionally, the `observation' Gaussian noise $\epsilon\sim\cN(0, \sigma^2)$
is incorporated into the GP model,
though it is used only for optimization stability with a negligibly small value $\sigma=10^{-6}$.
Without any prior knowledge, the prior mean model is assumed to be zero,
i.e. $\overline{\Xi}_{ij}(\bp)=0$.
The covariance kernel $K$ is defined using radial basis function (RBF) over the parameter space.
\kctwo{
The hyperparameter of the kernel $K$ is calibrated
to maximize the likelihood of the measurement data $\cD_{ij}$~\cite{rasmussen2003gaussian},
using the python \texttt{scikit} package~\cite{scikit-learn}.
}
Conditioned upon the data $\cD_{ij}$,
the probability of the coefficient at \refC{an} unseen parameter point, i.e. $\Xi_{ij}(\bp^*)$,
also follows a normal distribution~\cite{rasmussen2003gaussian},
\begin{equation}\label{eq:gp-predict} 
P(\Xi_{ij}(\bp^*) | \cD_{ij}, \theta_{GP})
\sim
\cN[\tilde{\Xi}_{ij}(\bp^*), \mathrm{Var}(\Xi_{ij}(\bp^*))],
\end{equation}
\kctwo{
where the expression for the conditional average and variance can be
found in~\cite{rasmussen2003gaussian}.
}
\par
A prediction at \refC{an} unseen parameter point $\bp^*$ would involve
either using the conditional average given in (\ref{eq:gp-predict})
or sampling from the probability (\ref{eq:gp-predict})
for all coefficients, to construct the dynamics coefficients $\bXi(\bp^*)$.
The estimated $\bXi(\bp^*)$ is then used to simulate the latent dynamics (\ref{eq:lasdi}),
with initial condition $\bz_0(\bp^*) = \En(\kc{T_0}(\bp^*))$
encoded from the initial state field for the parameter point $\bp^*$.
The latent space trajectory $\{\bz_1, \ldots, \bz_{N_t}\}$ is then
decoded back to the state field for prediction.
The overall framework is also illustrated in Figure~\ref{fig:schematic}.

\subsection{Greedy sampling}\label{sec:greedy}
If initial training data is scarce in parameter space
and not sufficient for effective parametric model training,
data at additional parameter points must be collected on the fly
in the training process.
In GPLaSDI framework,
the uncertainties of $\bXi(\bp^*)$ (\ref{eq:gp-predict})
\kctwo{can provide useful error metrics in the \refC{absence} of ground truth.}
\par
First, at each parameter point $\bp^*_i$,
$N_s$ samples of $\{\bXi_s(\bp^*_i)\}_{s=1}^{N_s}$ are collected from the probability distribution (\ref{eq:gp-predict})
\kctwo{in Monte-Carlo fashion}.
Samples of state field
are then collected from the simulations of the latent dynamics (\ref{eq:lasdi}) with the samples of $\bXi(\bp^*_i)$.
We denote this sample state field tensor as $\rT_{gp}\in\rR^{N_p^*\times N_t \times N_x \times N_s}$,
where $(\rT_{gp})_{ijkl} = \rT_{gp}(\bp^*_i, t_j, \bx_k, \bXi_l)$ represents
the HE state at grid point $\bx_k$, time $t_j$ for parameter $\bp^*_i$, predicted with the sample coefficients $\bXi_l$.
From $\rT_{gp}$, the solution uncertainty is evaluated with
the standard deviation of the sample state fields,
\begin{subequations}\label{eq:uq-metric}
\begin{equation}
\begin{split}
&\std_{gp}[\rT_{gp}](\bp^*_i) = \\
&\max_{t_j}
\frac{1}{N_x}\sum_{k=1}^{N_x}
\frac{1}{N_s}\sum_{s=1}^{N_s}
\{ \rT_{gp}(\bp^*_i, t_j, \bx_k, \bXi_s) - \overline{\rT}_{gp}(\bp^*_i, t_j, \bx_k) \}^2,
\end{split}
\end{equation}
where $\overline{\rT}_{gp}$ is the sample average of $\rT_{gp}$,
\begin{equation}
\overline{\rT}_{gp}(\bp^*_i, t_j, \bx_k) = \frac{1}{N_s}\sum_{s=1}^{N_s}\rT_{gp}(\bp^*_i, t_j, \bx_k, \bXi_s) \in \rR^{N_p^* \times N_x\times N_t}.
\end{equation}
\end{subequations}
The parameter point with the maximum uncertainty (\ref{eq:uq-metric}) is picked up as a new training parameter point,
\begin{equation}\label{eq:next-p}
\bp_{N_p+1} = \arg\max_i\;\std_{gp}[\rT_{gp}](\bp^*_i).
\end{equation}
The corresponding full-order model solution is then collected
and appended to the existing training data for the on-going training,
\begin{equation}
\rT = \{\bT(\bp_1), \bT(\bp_2), \ldots, \bT(\bp_{N_p}), \bT(\bp_{N_p+1})\}.
\end{equation}
The temporal average and standard deviation in (\ref{eq:T-scale0}) are updated accordingly.
We also refer readers to Bonneville \textit{et al.}~\cite{bonneville2024gplasdi} for the detailed algorithm of this greedy sampling process.

%% file: figures_schematic.tex
\begin{tikzpicture}[
    font=\scriptsize,
    ]
    
    \def\ldwidth{2.1}
    \def\ldheight{1.5}
    \def\ldgap{1.0}
    
    \draw[draw=black, fill=green!10, line width=1.0,] (0, 0) rectangle node[] (ld2) {$\dot{\bz} = \bPhi(\bz)\bXi(\bp_2)$} ++ (\ldwidth, \ldheight);
    \draw[draw=black, fill=green!10, line width=1.0,] (0, \ldheight + \ldgap) rectangle node[] (ld1) {$\dot{\bz} = \bPhi(\bz)\bXi(\bp_1)$} ++ (\ldwidth, \ldheight);
    \node[] (vdots) at (0.5 * \ldwidth, -0.5 * \ldgap) {$\vdots$};
    \draw[draw=black, fill=red!10, line width=1.0,] (0, -1.5 * \ldgap) rectangle node[] (uld) {$\dot{\bz} = \bPhi(\bz)\bXi(\bp)$} ++ (\ldwidth, -\ldheight);

    \node[inner sep=0pt, anchor=south, yshift=15pt, ] at (ld1.north) {\textbf{Observed dynamics 1}};
    \node[inner sep=0pt, anchor=south, yshift=15pt, ] at (ld2.north) {\textbf{Observed dynamics 2}};
    \node[inner sep=0pt, anchor=south, yshift=15pt, ] at (uld.north) {\textbf{Unseen dynamics}};

    \node[inner sep=0pt,] at (\ldwidth + 0.75, 0.5 * \ldheight) {$\Longrightarrow$};
    \node[inner sep=0pt,] at (- 0.75, 0.5 * \ldheight) {$\Longrightarrow$};

    \node [trapezium, trapezium angle=60, minimum width=3cm, draw, rotate=90, line width=1.0, anchor=north west, fill=orange!10,] (decoder) at (\ldwidth + 1.5, 0.1) {};
    \node [trapezium, trapezium angle=60, minimum width=3cm, draw, rotate=-90, line width=1.0, anchor=north east, fill=orange!10,] (encoder) at (- 1.5, 0.1) {};

    \node[inner sep=0pt, anchor=west, xshift=10pt, ] at (decoder.south) {$\Longrightarrow$};
    \node[inner sep=0pt, anchor=east, xshift=-10pt, ] at (encoder.south) {$\Longrightarrow$};

    \node[inner sep=0pt, anchor=west, xshift=30pt] (pred) at (decoder.south)
    {\includegraphics[width=.2\textwidth]{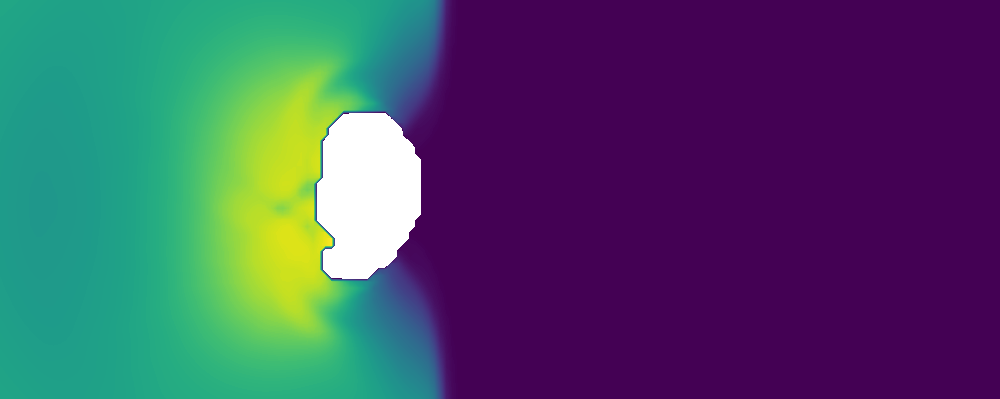}};
    \node[inner sep=0pt, anchor=east, xshift=-30pt] (truth) at (encoder.south)
    {\includegraphics[width=.2\textwidth]{data_Xtrain36.pure_zone.case35.timestep35.png}};

    \node[inner sep=0pt, anchor=south, xshift=0pt, yshift=2pt,] at (pred.north) {\textbf{Prediction} $\bT_{pred}(\bp)$};
    \node[inner sep=0pt, anchor=south, xshift=0pt, yshift=2pt,] at (truth.north) {\textbf{Ground truth} $\bT(\bp)$};

    \node[inner sep=0pt, anchor=south, yshift=12pt,] (decodereq) at (decoder.east) {$\THE=\De(\bz;\theta)$};
    \node[inner sep=0pt, anchor=south, yshift=5pt,] at (decodereq.north) {\textbf{Decoder}};

    \node[inner sep=0pt, anchor=south, yshift=12pt,] (encodereq) at (encoder.west) {$\bz=\En(\THE;\theta)$};
    \node[inner sep=0pt, anchor=south, yshift=5pt,] at (encodereq.north) {\textbf{Encoder}};
    
    \end{tikzpicture}

%% file: sections_ae_jcp.tex
\subsection{The proposed interface-tracking autoencoder}\label{sec:ae}



We introduce an indicator variable to distinguish the physical domain from the void or inactive region separated by a sharp interface:
\begin{equation}
\bphi(T(\bx; \bp)) = \mathbf{1}_{T>0}(\bx; \bp),
\end{equation}
where $\bphi$ equals 1 in the active (non-pore) domain and 0 in the pore region.
\par
The interface tracking autoencoder is designed to simultaneously capture the
state field and the pore domain using this indicator variable.
Specifically, the decoder now decodes both the (scaled) state field and the
indicator variable from the latent space variable,
\begin{equation}\label{eq:de1}
(\tilde{T}', \bphi_{pred}) = \De(\bz; \theta_{de}),
\end{equation}
\kc{
where the last output layer width is doubled to accommodate both outputs:
$d_{L+1} = 2N_x$.
While a larger MLP architecture size could be used for the decoder,
the same MLP architecture in Section~\ref{sec:standard-ae} was sufficient in this study.
}
To ensure the decoded indicator $\bphi_{pred}$ to range from 0 to 1,
a sigmoid activation function $\sigma_{sig}$ was applied to the last layer,
\begin{equation}\label{eq:lastlayer}
\begin{pmatrix}
\tilde{T}'\\ \bphi_{pred}
\end{pmatrix}
\equiv \cT_{L+1}(\bq_L)
=
\begin{pmatrix}
\bW_{L+1}\bq_L + \bb_{L+1}\\ \sigma_{sig}(\bW'_{L+1}\bq_L + \bb'_{L+1})
\end{pmatrix}.
\end{equation}
Here, the decoded state field $\tilde{T}'$ is defined over the entire spatial
domain and does not explicitly delineate the active region. In contrast, the
decoded indicator $\bphi_{\text{pred}} \in [0, 1]^{N_x}$ captures the spatial
extent of the material domain and effectively serves as a probability-like
measure, indicating the likelihood that each grid point belongs to the active
(non-pore) region.
The final reconstructed state field $\tilde{T}_{pred}$
is computed combining $\tilde{T}'$ and $\bphi_{pred}$,
\begin{equation}\label{eq:Tpred1}
\tilde{T}_{pred}(\bx; \bp) = \tilde{T}'(\bx; \bp) \times \mathbf{1}_{\bphi_{pred}>0.5}(\bx; \bp).
\end{equation}
This formulation ensures that the reconstructed state field \refC{is} zero in regions
where $\bphi_{pred}<=0.5$, regardless of the value of $\tilde{T}'$.  Therefore,
if $\bphi_{pred}$ is reconstructed accurately, the moving interface is tracked
correctly.
\par
\begin{figure*}[htbp]
    \input{figures_ae_architecture.tex}
    \caption{Schematic of the revised autoencoder architecture.\KC{Techinically, we only expand only output layer as twice, and hidden layers remain the same as encoder.}}
    \label{fig:ae-architecture}
\end{figure*}
While the encoder could also be modified to take $\bphi$ as an input as well,
we take the rationale that the autoencoder must be capable of extracting $\bphi$ out of $T$,
since $\bphi$ is essentially a filtered information of $T$.
\kc{
Experiments showed that the encoder performed better when taking the state
field $T$ solely as an input.
}
\todo{Add these results as appendix?}
The architecture of the revised autoencoder for LaSDI-IT is illustrated in
Figure~\ref{fig:ae-architecture}.

\subsubsection{Autoencoder loss}
We denote as $\bphi_k(\bp_i)$ the indicator corresponding
to the training data state field $T_k(\bp_i)$ at time step $k$ for the parameter $\bp_i$.
The reconstruction of training data $\tilde{T}_{k,pred}(\bp_i)$
and $\bphi_{k, pred}(\bp_i)$ are evaluated as
\begin{equation}
\begin{pmatrix}
\tilde{T}_{k,pred}(\bp_i) \\
\bphi_{k, pred}(\bp_i)
\end{pmatrix}
=
\De(\En(T_k(\bp_i); \theta_{en}); \theta_{de}).
\end{equation}
The autoencoder loss (\ref{eq:J-ae0}) is now revised for LaSDI-IT to track both
$\tilde{T}_{pred}$ and $\bphi_{pred}$,
\begin{subequations}\label{eq:J-ae1}
\begin{equation}
\cJ_{AE} = \cJ_{val} + \cJ_{mask},
\end{equation}
which is composed of value loss $\cJ_{val}$ and mask loss $\cJ_{mask}$,
\begin{equation}\label{eq:J-value}
\cJ_{val} = \frac{1}{N_pN_tN_x}\sum_i^{N_p}\sum_k^{N_t}
\Vert \tT_k(\bp_i) - \tilde{T}_{k,pred}(\bp_i) \Vert^2
\end{equation}
\begin{equation}\label{eq:J-mask}
\cJ_{mask} = \frac{1}{N_pN_tN_x}\sum_i^{N_p}\sum_k^{N_t}
\Vert \bphi_k(\bp_i) - \bphi_{k, pred}(\bp_i) \Vert^2.
\end{equation}
\end{subequations}

\subsubsection{Revised scaling of the state field}
Since the states in the pore region are not physically meaningful but serve
only as placeholders for the indicator, the temporal average and standard
deviation in (\ref{eq:T-scale0}) should exclude these values (e.g., $0$ for
pore-collapse application) within the pore region.  Thus we count the sample
size of active state at each grid point of the training data,
\begin{equation}
N_{active}(\bx_g) = \frac{1}{N_pN_t}\sum_{i=1}^{N_p}\sum_{k=1}^{N_t}\bphi_k(\bx_g; \bp_i) \in \rR^{N_x},
\end{equation}
and evaluate average and standard deviation only for non-zero active state values,
\begin{subequations}\label{eq:T-scale1}
\begin{equation}
\overline{T}(\bx_g) = \frac{1}{N_{active}(\bx_g)}\frac{1}{N_pN_t}\sum_{i=1}^{N_p}\sum_{k=1}^{N_t}T_k(\bx_g; \bp_i)\bphi_k(\bx_g; \bp_i)
\end{equation}
\begin{equation}
\begin{split}
\std[T](\bx_g) = &\left\{ \frac{1}{N_{active}(\bx_g)}\frac{1}{N_pN_t}\sum_{i=1}^{N_p}\sum_{k=1}^{N_t} \right.\\
&\qquad (T_k(\bx_g; \bp_i)\bphi_k(\bx_g; \bp_i) - \overline{T}(\bx_g))^2 \bigg\}^{1/2}.
\end{split}
\end{equation}
\end{subequations}
\kctwo{
The standard scaling (\ref{eq:T-scale}) is then used with the newly defined average and standard deviation in (\ref{eq:T-scale1}).
}

%% file: figures_ae_architecture.tex
\begin{tikzpicture}[
    font=\scriptsize,
    ]
    
    \def\ldwidth{2.1}
    \def\ldheight{1.5}
    \def\ldgap{1.0}
    



    \node [trapezium, trapezium left angle=20, trapezium right angle=60, minimum width=5cm, draw, rotate=90, line width=1.0, anchor=north west, fill=orange!10,] (decoder) at ( 0.5, 0.1) {};
    \node [trapezium, trapezium angle=60, minimum width=2.5cm, draw, rotate=-90, line width=1.0, anchor=north east, fill=orange!10,] (encoder) at ( -0.5, 0.1) {};
    \draw[draw=purple, line width=6.,] (1.7, -3.15) node[anchor=north east, xshift=.5cm,] {Sigmoid} -- ++(0, 2.5cm);

    \draw[->, line width=1.] (decoder.south) + (5pt, 0pt) -- ++ (20pt, 0pt)
        node[inner sep=0pt, anchor=west, xshift=5pt,] (T1)
        {\includegraphics[width=.2\textwidth]{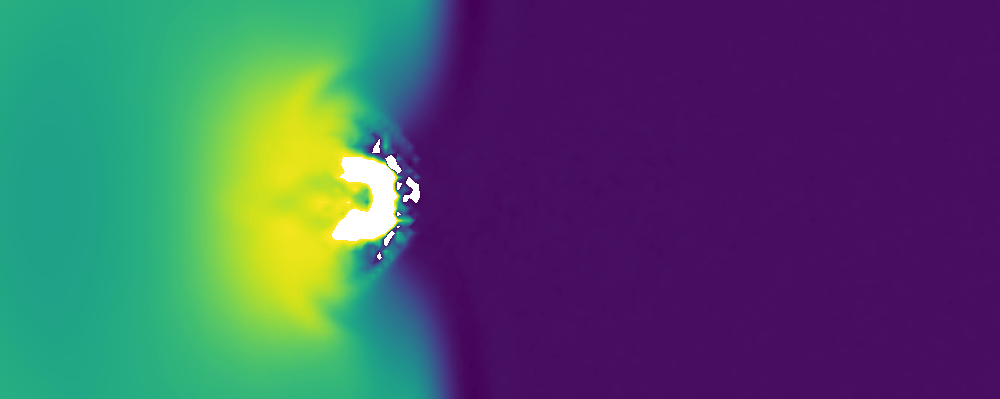}};
    \draw[->, line width=1.] (decoder.south) + (5pt, -2.5cm) -- ++ (20pt, -2.5cm)
        node[inner sep=0pt, anchor=west, xshift=5pt,] (mask)
        {\includegraphics[width=.2\textwidth]{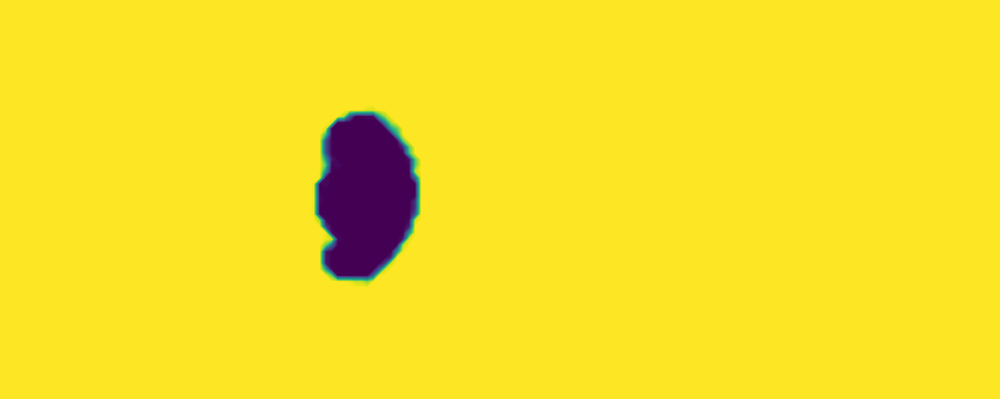}};

    \draw[->, line width=1.] (T1.east) + (5pt, 0pt) -- ++ (20pt, -20pt)
        node[inner sep=0pt, anchor=west, xshift=5pt, yshift=-15pt,] (pred)
        {\includegraphics[width=.2\textwidth]{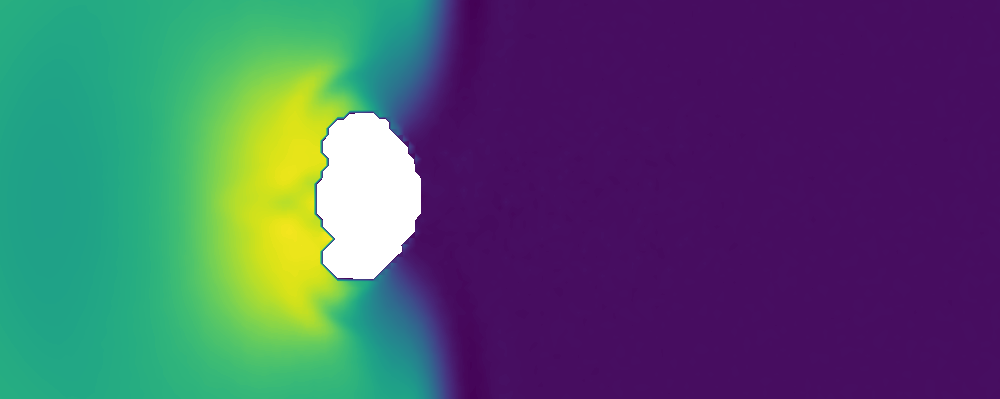}};
    \draw[->, line width=1.] (mask.east) + (5pt, 0pt) -- ++ (20pt, 20pt);


    \draw[<-, line width=1.] (encoder.south) + (-5pt, 0pt) -- ++ (-20pt, 0pt)
        node[inner sep=0pt, anchor=east, xshift=-5pt, yshift=0pt,] (truth)
        {\includegraphics[width=.2\textwidth]{data_Xtrain36.pure_zone.case35.timestep35.png}};

    \draw[->, dashed, line width=1.] (truth.south) + (0pt, -5pt) -- ++ (0pt, -20pt)
        node[inner sep=0pt, anchor=north, xshift=0pt, yshift=-15pt,] (truthmask)
        {\includegraphics[width=.2\textwidth]{data_Xtrain36.case17.timestep35.mask.png}};

    \node[inner sep=0pt, anchor=south, xshift=0pt, yshift=2pt,] at (truth.north) {$\THE\in\rR^{N_x}$};
    \node[inner sep=0pt, anchor=south, xshift=0pt, yshift=2pt,] at (T1.north) {$\tilde{T}'\in\rR^{N_x}$};
    \node[inner sep=0pt, anchor=south, xshift=0pt, yshift=2pt,] at (mask.north) {$\phi_{pred}\in\rR^{N_x}$};
    \node[inner sep=0pt, anchor=south, xshift=0pt, yshift=2pt,] at (pred.north) {$\tilde{T}_{pred}=\tilde{T}'\times\mathbf{1}_{\phi_{pred}>0.5}$};
    \node[inner sep=0pt, anchor=south, xshift=0pt, yshift=2pt,] at (truthmask.north) {$\phi=\mathbf{1}_{T>0}\in\rR^{N_x}$};

    \node[inner sep=0pt, anchor=south, yshift=12pt,] (decodereq) at (decoder.east) {$(\tilde{T}', \phi)=\De(\bz;\theta)$};
    \node[inner sep=0pt, anchor=south, yshift=5pt,] at (decodereq.north) {\textbf{Decoder}};

    \node[inner sep=0pt, anchor=south, yshift=12pt,] (encodereq) at (encoder.west) {$\bz=\En(\tilde{T};\theta)$};
    \node[inner sep=0pt, anchor=south, yshift=5pt,] at (encodereq.north) {\textbf{Encoder}};
    
    \end{tikzpicture}

%% file: sections_results_jcp.tex
\section{Results}\label{sec:results}
While the vanilla LaSDI framework includes an $L_2$-norm regularization term
(i.e., $\beta_2$ in Eq.~\ref{eq:J}) for the coefficient matrix $\bXi$, we found
it unnecessary for the pore collapse problem. Therefore, we set $\beta_2 = 0$
for all numerical experiments.
\par
\input{sections_limitation}
\par
\input{sections_reconstruction}
\par
\input{sections_scaling}

\par
\subsection{The performance of LaSDI-IT}
The entire LaSDI-IT framework in Section~\ref{sec:lasdi} is then applied to identify latent dynamics of
single pore collapse processes.
This section demonstrates the effectiveness of the GPLaSDI framework, particularly in data-scarce scenarios.
\par
While SINDy supports general nonlinear
formulations, we restrict the latent dynamics to a linear form, motivated by
two factors.  First, limiting to linear dynamics significantly reduces the
number of coefficients to be fitted, decreasing the risk of overfitting without
sparse regression.  This allows us to prevent a potential slow-down in
multi-objective training with an additional sparsity regularization term.
Second, while the state field exhibits complex features, its evolution is
predominantly one-directional as the shock propagates through the domain.  Once
the shock passes and the pore fully collapses, the state field reaches a
quasi-steady state.  This observation leads to an expectation that the
evolution of the state field can be sufficiently represented by linear dynamics
with a single fixed point with the autoencoder capturing the nonlinear solution
manifold.
\par
For physics parameters,
the angle $p_1$ and the length $p_2$ of the major axis of the pore are considered
for ranges of $\bp = (p_1, p_2) \in [\deg{0}, \deg{20}]\times [1\mu m, 1.2\mu m]$.
To simulate a data-scarce situation,
the initial training dataset consists of only four parameter points
located at the corners of the parameter space:
$\rP^{(0)} = \{(\deg{0}, 1\mu m), (\deg{20}, 1\mu m), (\deg{0}, 1.2\mu m), (\deg{20}, 1.2\mu m)\}$.
For an inner-loop training,
the revised autoencoder is trained on the available training data
to minimize the objective (\ref{eq:J}) with $\beta_1 = 10^2\Delta t^2$,
where $\cJ_{AE}$ and $\cJ_{LD}$ are defined in (\ref{eq:J-ae1}) and (\ref{eq:J-ld}), respectively.
Each inner-loop training took $2000$ iterations of ADAM optimization
with a learning rate of $10^{-4}$.
\kctwo{
For the greedy sampling in Section~\ref{sec:greedy},
the solution uncertainty (\ref{eq:uq-metric}) is evaluated
on $11\times11$ test points uniformly distributed on $[\deg{0}, \deg{20}]\times [1\mu m, 1.2\mu m]$,
each with 30 Monte-Carlo samples.
Once a new parameter point was added to the training dataset per (\ref{eq:next-p}),
the outer-loop process was repeated with gradually increasing training data.
}
\par
\kc{
Note that, for the uncertainty metric (\ref{eq:uq-metric}), we
averaged the sample standard deviation over space as uncertainty metric,
while Bonneville \textit{et al.}~\cite{bonneville2024gplasdi}
takes the maximum sample standard deviation over both space and time.
Considering the temperature field data shown in Figure~\ref{fig:he-temp}, sample
variance for masking pore region causes high-temperature standard deviation near the
discontinuous pore interface.  Maximum sample standard deviation will be always
chosen from the pore interface region, thus not appropriately representing the
uncertainty of the overall solution.  For this reason, we evaluate the sample
standard deviation to be averaged over space.
}
\par
\kc{
The training is performed in parallel with multiple NVidia V100 GPUs on LC Lassen~\cite{lassen}.
For inner loop training, each dataset at a training parameter point is assigned to one GPU process,
starting from 4 parameter points (4 GPUs) until the dataset reaches 15 parameter points (15 GPUs).
At each outer loop iteration,
GP-based uncertainty quantification and sampling the new training parameter point
are performed on a IBM Power9 CPU.
Throughout the training, the inner loop takes about 3.9 minutes per dataset
while the outer loop takes about 1.3 minutes.
The overall training takes about 8.0 hours of computation time,
which is in fact shorter than that of one high-fidelity simulation at a parameter point.
}
\par
\begin{figure}[tbhp]
    \input{figures_greedy_sampling.tex}
    \caption{Gaussian-process based greedy sampling:
    (a) solution uncertainty (\ref{eq:uq-metric}) and
    (b) relative error (\ref{eq:rel-error}) compared to the ground truth at the 14-th iteration;
    (c) relative error (\ref{eq:rel-error}) at the 15-th iteration with an additionally sampled training data point;
    and (d) relative error (\ref{eq:rel-error}) from uniformly sampled initial data as the reference.
    The training data at the current outer-loop iteration is marked with a black box,
    and the initial training data is marked with a red box.}
    \label{fig:greedy-sampling}
\end{figure}
The propagated solution uncertainty (\ref{eq:uq-metric})
is visualized in Figure~\ref{fig:greedy-sampling}~(a).
Figure~\ref{fig:greedy-sampling}~(b) shows the relative error of LaSDI prediction
compared to the ground truth,
\begin{equation}\label{eq:rel-error}
\epsilon[\bT(\bp)]
=
\max_k\frac{\Vert T_k(\bp) - T_{k, pred}(\bp) \Vert_2}{\Vert T_k(\bp) \Vert_2},
\end{equation}
where the maximum value is taken over all timesteps.
We emphasize that, in the absence of the ground truth data,
this relative error is unavailable.
The solution uncertainty (\ref{eq:uq-metric})
exhibits a remarkably strong correlation with the relative error,
though evaluated without the ground truth.
\par
This demonstrates the utility of the solution uncertainty as an error metric for guiding greedy sampling.
Figure~\ref{fig:greedy-sampling}~(c) shows the relative error
after the next 15th outer-loop iteration,
where the new training data point $(p_1, p_2) = (\deg{0}, 1.16\mu m)$
is added based on the solution uncertainty from the 14th outer loop iteration (Figure~\ref{fig:greedy-sampling}a).
Compared to the previous iteration (Figure~\ref{fig:greedy-sampling}b),
the overall prediction accuracy improved dramatically,
achieving a relative error of $\lesssim9\%$ in all test cases.
For comparison,
a reference case was trained using a uniformly sampled dataset of 36 parameter points arranged as a 
$6\times6$ grid over the parameter space $[\deg{0}, \deg{20}]\times [1\mu m, 1.2\mu m]$.
This reference case did not employ greedy sampling, and the inner-loop training was performed for 
$3\times10^4$ iterations---equivalent to the total number of inner-loop iterations in the greedy sampling case.
As shown in Figure~\ref{fig:greedy-sampling}~(d),
the prediction from this uniform sampling achieved a relative error of $\lesssim 8\%$ across all test cases.
Remarkably, the GP-based greedy sampling approach achieved comparable accuracy
with only half the amount of training data, underscoring its efficiency in data-scarce scenarios.
\par
\begin{figure}[tbph]
    \input{figures_metrics.tex}
    \caption{The worst case prediction $(\deg{12}, 1.2\mu m)$ from greedy-sampling training:
    (a) hot spot area ($T>800K$) \eqref{eq:A-hotspot};
    (b) pore area \eqref{eq:A-pore};
    and (c) maximum temperature \eqref{eq:max-temp}.}
    \label{fig:metrics}
\end{figure}
Figure~\ref{fig:metrics} shows \kctwo{three key metrics (Eq.~\ref{eq:A-hotspot}--\ref{eq:max-temp})}
for the worst case prediction from the greedy sampling case,
corresponding to $(p_1, p_2) = (\deg{12}, 1.2\mu m)$,
which achieved a relative error of approximately $9.3\%$.
The predicted hot spot area and pore area closely match the ground truth,
while the peak of the maximum temperature is slightly underestimated.
This underestimation is likely due to the highly localized and transient nature of the temperature peak,
which may be smoothed during the embedding process in the latent space.
Nevertheless, the model accurately predicts the timing of the peak temperature and the subsequent evolution,
which are critical for upscaling HE material properties.
\kc{
It is worth emphasizing that
latent dynamics prediction of one parameter point takes only about $0.033s$
with \refC{an IBM Power9} CPU processor~\cite{lassen}.
\refC{This} computational time is about $2.0\times10^6$ times faster than the high-fidelity simulation in Section~\ref{sec:porecollapse}.
}
\par
\begin{figure}[tbph]
    \input{figures_prediction.tex}
    \caption{Ground truth (left), prediction (middle), and the error (right) of the worst case prediction $(\deg{12}, 1.2\mu m)$ from greedy-sampling training at
    (a) $t=0.1ns$
    (b) $t=0.2ns$
    (c) $t=0.3ns$
    (d) $t=0.4ns$
    (e) $t=0.5ns$
    (f) $t=0.6ns$
    and (g) $t=1.2ns$.
    Errors due to pore mask mis-prediction are marked as white (right).}
    \label{fig:prediction}
\end{figure}
Figure~\ref{fig:prediction}
compares the predicted and ground truth temperature fields for the worst-case prediction at various timesteps.
The latent dynamics model qualitatively captures the evolution of the pore shape and the hot spot after pore collapse.
Aside from pore shape mis-prediction, errors are primarily localized near the shock front
and the boundaries of the hot spots, where physical discontinuities in temperature are present.
These errors are attributed to the spectral bias of the autoencoder,
though these physical discontinuities are much milder than those at the pore boundary.
The maximum point-wise error reaches approximately \kc{$193K$}
as the pore collapses and hot spot formation,
but decreases down to about $30K$ as the hot spot dissipates.
Despite these localized errors,
the latent dynamics model effectively predicts both the detailed instantaneous evolution
of the temperature and the pore, and the key metrics required for HE material design.

%% file: sections_limitation.tex
\subsection{Limitations with the standard autoencoder}\label{subsec:vanilla}
We assess the performance of the standard autoencoder in the vanilla LaSDI framework,
introduced in Section~\ref{sec:standard-ae}, on the interface-tracking problem of
shock-induced pore collapse. The encoder network (\ref{eq:en}) is
implemented as a MLP (\ref{eq:mlp}) with $L = 6$ hidden layers and
layer widths $(d_0, d_1, \ldots, d_7) = (N_x, 3000, 300, 300, 100, 100, 30,
N_z)$. The decoder (\ref{eq:de}) mirrors this architecture with the same
layer widths in reverse order.
\par
To evaluate the reconstruction capability of the autoencoder independently of
latent dynamics modeling, we set $\beta_1 = 0$ in the loss function
(\ref{eq:J}), thereby focusing solely on reconstruction loss of the
autoencoder. Training is performed using the ADAM
optimizer~\cite{kingma2014adam} for $10^4$ iterations with a learning rate of
$10^{-4}$. The training dataset includes $N_p = 4$ parameter configurations:
$\{(\deg{0}, 1\,\mu\text{m}), (\deg{20}, 1\,\mu\text{m}), (\deg{0},
1.2\,\mu\text{m}), (\deg{20}, 1.2\,\mu\text{m})\}$.
\par
\begin{figure}[tbph]
    \input{figures_vanilla_ae.tex}
    \caption{(a) Reconstruction of the state field with the standard AE, and (b) its point-wise error.}
    \label{fig:vanilla-ae}
\end{figure}
Despite its simplicity, this standard autoencoder architecture struggles to
accurately encode the geometry of the pore into the latent space.
Figure~\ref{fig:vanilla-ae} presents the reconstructed HE temperature field
using a trained autoencoder with latent dimension $N_z = 10$. As shown in
Figure~\ref{fig:vanilla-ae}(a), the reconstruction captures the overall
features of the temperature field reasonably well across most of the domain.
However, the reconstruction error in Figure~\ref{fig:vanilla-ae}(b) reveals
significant discrepancies concentrated near the pore interface, indicating a
failure to resolve the sharp material boundary.
\par
Additionally, the reconstructed pore region contains nonzero values, which are
physically invalid. While some of these \kctwo{temperature values are obviously unrealistic,
suggesting pore presence,} others lie near the reference value,
making the pore boundary ambiguous. This ambiguity compromises accurate
estimation of the pore area---an essential metric in high-explosive (HE) material
characterization and design. Alternative network configurations, including
increased depth, width, and the use of convolutional layers, were explored, but
these modifications did not resolve the interface reconstruction issue.
\par
As discussed earlier, the primary difficulty arises from the sharp
discontinuity in the state field across the pore boundary, as illustrated in
Figure~\ref{fig:he-temp}. From a spectral perspective, accurately resolving
such discontinuities requires capturing a wide range of spatial wavenumber
modes. However, neural networks are known to exhibit \emph{spectral
bias}~\cite{rahaman2019spectral}, favoring the learning of low-frequency
components over high-frequency ones. As a result, standard autoencoders
struggle to capture high-frequency features such as steep gradients or
discontinuities at material interfaces. While increasing the network depth and
width may improve resolution, it comes at the cost of significantly longer
training times and potentially more difficult optimization.

%% file: figures_vanilla_ae.tex
\begin{tikzpicture}[
    font=\small,
    spy using outlines={circle,black,magnification=6,size=1.5cm, connect spies}
    ]
    
    \pgfplotsset{set layers=standard}
    \begin{groupplot}[
        group style={
            group name = my plots,
            group size= 2 by 1,
            xlabels at =edge bottom,
            horizontal sep=3.cm,
            vertical sep=2.cm,
        },
        name=chung,
    ]    

        \nextgroupplot[
            height = 0.25\textwidth,
            width = 0.45\textwidth,
            enlarge x limits={false, abs value = 0mm},
            ylabel={$x_2$},
            xlabel={$x_1$},
            tick scale binop ={\times},
            xmin=0e-4,xmax=7e-4,ymin=0e-4,ymax=3e-4,
            point meta min=-1., point meta max=809,
            colorbar, colormap/viridis,
            colorbar style={
                font=\scriptsize,
                xticklabel pos=upper,
                scaled y ticks=false,
                /pgf/number format/precision=4,
                xlabel=$T_{pred}$,
            }
        ]
        
            \edef\imagepath{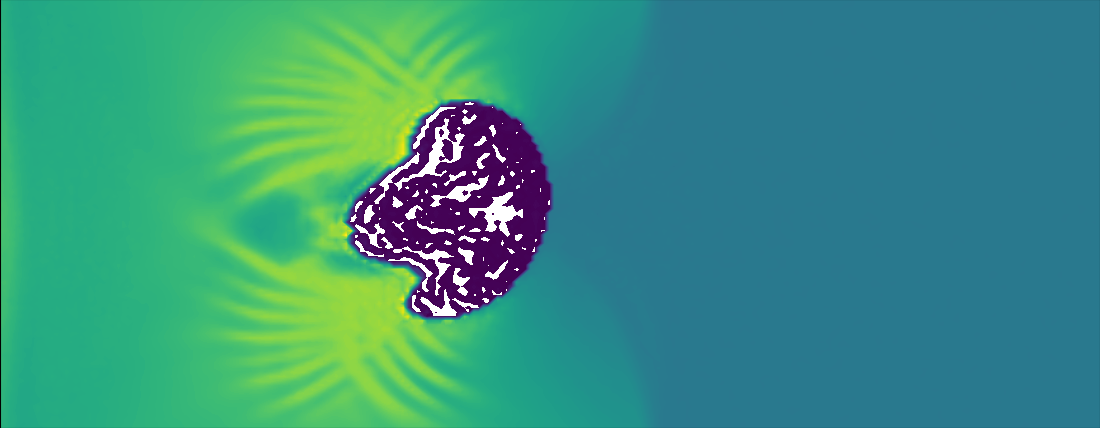}
            \addplot graphics[xmin=0e-4,xmax=7e-4,ymin=0e-4,ymax=3e-4]{\imagepath};

        \nextgroupplot[
            height = 0.25\textwidth,
            width = 0.45\textwidth,
            enlarge x limits={false, abs value = 0mm},
            ylabel={$x_2$},
            xlabel={$x_1$},
            tick scale binop ={\times},
            xmin=0e-4,xmax=7e-4,ymin=0e-4,ymax=3e-4,
            point meta min=0., point meta max=201,
            colorbar, colormap/viridis,
            colorbar style={
                font=\scriptsize,
                xticklabel pos=upper,
                scaled y ticks=false,
                /pgf/number format/precision=4,
                xlabel=$\vert T_{pred}-T\vert$,
            }
        ]
        
            \edef\imagepath{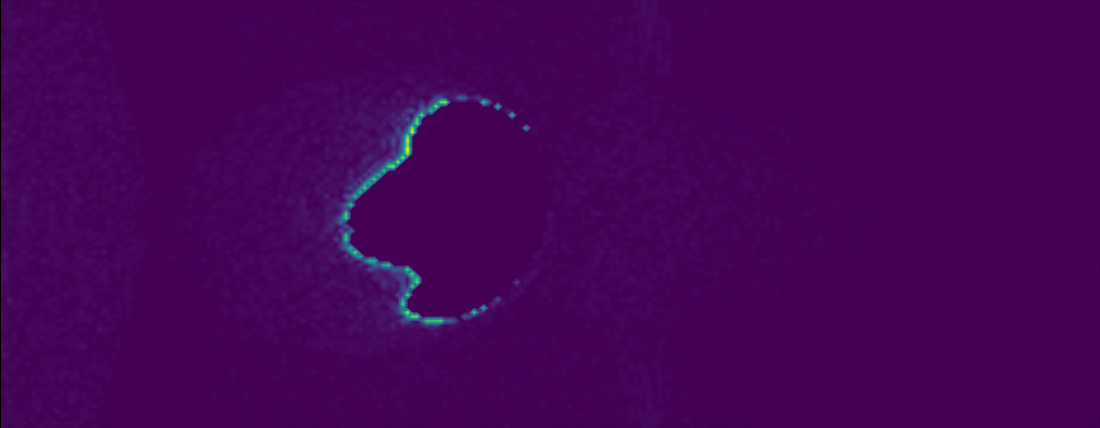}
            \addplot graphics[xmin=0e-4,xmax=7e-4,ymin=0e-4,ymax=3e-4]{\imagepath};
    
    \end{groupplot}
    \node[below = 1.5cm of my plots c1r1.south west,
        anchor=west,
    ] {(a) $T_{pred}$};
    \node[below = 1.5cm of my plots c2r1.south west,
        anchor=west,
    ] {(b) Point-wise error};
\end{tikzpicture}
    

%% file: sections_reconstruction.tex
\subsection{Reconstruction results with the autoencoder in the LaSDI-IT}
To enable a direct comparison with the standard autoencoder,
the newly proposed autoencoder is trained using the same dataset of
$N_p = 4$ parameter points described in Section~\ref{subsec:vanilla}.
\kctwo{
The same standard MLP architecture in Section~\ref{subsec:vanilla}
is used for the encoder network (\ref{eq:en}).
}
The decoder (\ref{eq:de1}) mirrors this configuration in
reverse, except for the final output dimension $d_7 = 2N_x$ and the application
of a sigmoid activation function in the final layer, as described in
\eqref{eq:lastlayer}. The revised scaling approach (\ref{eq:T-scale1})
is applied, and the modified AE loss (\ref{eq:J-ae1}) is minimized
using the ADAM optimizer with a learning rate of $10^{-4}$ over $10^4$
iterations. Both encoder and decoder maintain the same hidden layer structure
as used in the \kctwo{standard AE}.
\par
\begin{figure}[tbh]
    \input{figures_ae_loss.tex}
    \caption{Loss history of the standard AE and revised AE.}
    \label{fig:ae-loss}
\end{figure}
Figure~\ref{fig:ae-loss} shows the training loss history of the interface-tracking
AE compared to the standard version from Section~\ref{subsec:vanilla}.
The mask loss (\ref{eq:J-mask}) decreases by roughly four orders of
magnitude, demonstrating the model’s ability to accurately identify the HE
domain. Similarly, the value loss (\ref{eq:J-value}) converges rapidly,
significantly faster than the reconstruction loss
(\ref{eq:J-ae0}) of the standard AE.
\begin{figure}[tbh]
    \input{figures_revised_ae.tex}
    \caption{(a) Reconstruction of the temperature field with the AE in
    LaSDI-IT, and (b) its point-wise error.}
    \label{fig:revised-ae}
\end{figure}
Figure~\ref{fig:revised-ae} shows the reconstructed temperature field produced
by the interface-tracking AE. In contrast to the standard AE, the
reconstruction exhibits minimal error near the pore boundary, with no spurious
high-temperature artifacts. Moreover, the pore region is clearly defined by
distinct $0\,\text{K}$-value region, eliminating ambiguity in identifying the pore
boundary.
\par
Although the reconstruction loss for LaSDI-IT (\ref{eq:J-ae1}) involves
multi-objective minimization, it yields significantly better results than the
standard single-objective formulation. This improvement can be attributed to
two key factors. First, minimizing the mask loss (\ref{eq:J-mask})
effectively poses a classification task at each grid point, which is generally
easier than regressing to exact $0\,\text{K}$ values in the pore region.
Second, by excluding the $0\,\text{K}$ values from the statistics,
the temperature variance within the pore region is substantially reduced. As a
result, the autoencoder---by jointly learning both temperature and indicator
fields---focuses on reconstructing only the HE temperature distribution, which
exhibits lower variance and is more amenable to accurate regression.

%% file: figures_ae_loss.tex
\begin{tikzpicture}[
    font=\small,
    ]
    \begin{groupplot}[
        group style={
            group name = my plots,
            group size= 1 by 1,
            xlabels at =edge bottom,
            horizontal sep=2.5cm,
            vertical sep=2.2cm,
        },
        name=chung,
    ]    

        \nextgroupplot[
            height = 0.45\textwidth,
            width = 0.5\textwidth,
            enlarge x limits={false, abs value = 0mm},
            ylabel={Loss},
            xlabel={Iterations},
            tick scale binop ={\times},
            xmode=log, ymode=log,
            xmin=1, 
            ymin=1.0535157219436104e-05,
            ymax=1.9393409451113344,
            legend pos=south west,
            legend style={
                font=\scriptsize,
                draw=none, fill=none,
                legend cell align={left},
            },
        ]
        
        \addlegendimage{solid, black, line width=1.};
        \addlegendentry{Standard AE}
        \addlegendimage{solid, blue, line width=1.};
        \addlegendentry{Revised AE, value loss}
        \addlegendimage{solid, red, line width=1.};
        \addlegendentry{Revised AE, mask loss}

        \edef\imagepath{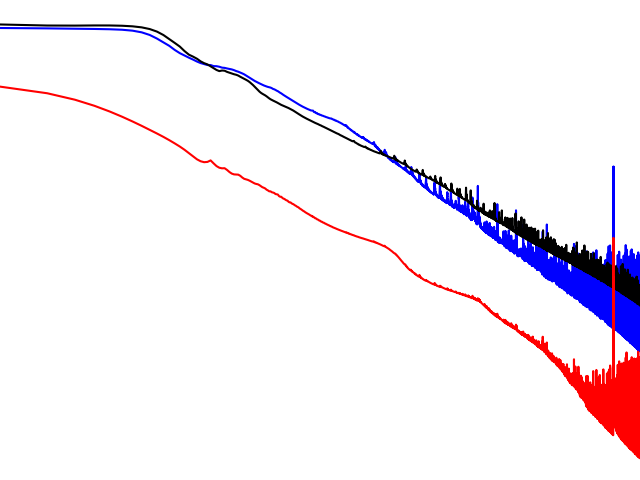}
        \addplot graphics[xmin=1,xmax=12000,
                            ymin=1.0535157219436104e-05,
                            ymax=1.9393409451113344]{\imagepath};
    
    \end{groupplot}
\end{tikzpicture}
    

%% file: figures_revised_ae.tex
\begin{tikzpicture}[
    font=\small,
    spy using outlines={circle,black,magnification=6,size=1.5cm, connect spies}
    ]
    
    \pgfplotsset{set layers=standard}
    \begin{groupplot}[
        group style={
            group name = my plots,
            group size= 2 by 1,
            xlabels at =edge bottom,
            horizontal sep=3.cm,
            vertical sep=2.cm,
        },
        name=chung,
    ]    

        \nextgroupplot[
            height = 0.25\textwidth,
            width = 0.45\textwidth,
            enlarge x limits={false, abs value = 0mm},
            ylabel={$x_2$},
            xlabel={$x_1$},
            tick scale binop ={\times},
            xmin=0e-4,xmax=7e-4,ymin=0e-4,ymax=3e-4,
            point meta min=317., point meta max=776,
            colorbar, colormap/viridis,
            colorbar style={
                ytick={317, 400, 600, 776},
                yticklabels={317, 400, 600, 776},
                font=\scriptsize,
                xticklabel pos=upper,
                scaled y ticks=false,
                /pgf/number format/precision=4,
                xlabel=$T_{pred}$,
            }
        ]
        
            \edef\imagepath{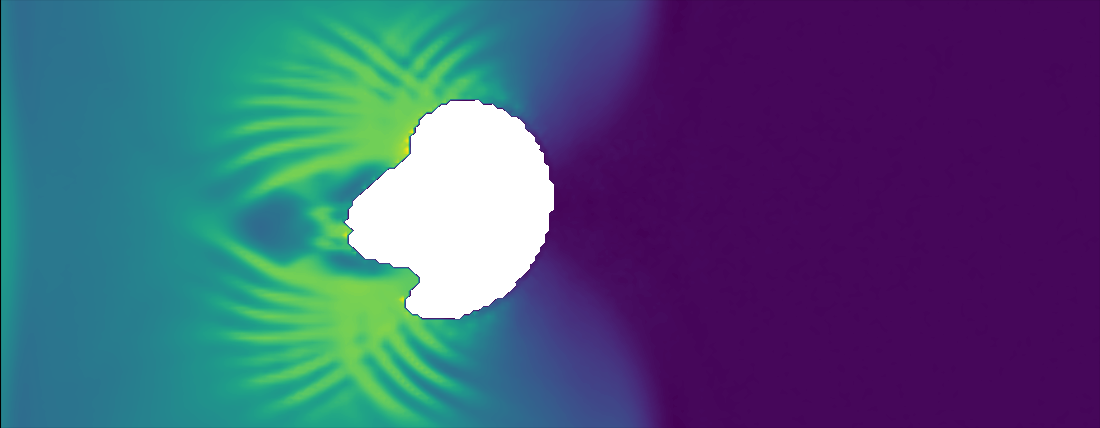}
            \addplot graphics[xmin=0e-4,xmax=7e-4,ymin=0e-4,ymax=3e-4]{\imagepath};

        \nextgroupplot[
            height = 0.25\textwidth,
            width = 0.45\textwidth,
            enlarge x limits={false, abs value = 0mm},
            ylabel={$x_2$},
            xlabel={$x_1$},
            tick scale binop ={\times},
            xmin=0e-4,xmax=7e-4,ymin=0e-4,ymax=3e-4,
            point meta min=0., point meta max=63,
            colorbar, colormap/viridis,
            colorbar style={
                font=\scriptsize,
                xticklabel pos=upper,
                scaled y ticks=false,
                /pgf/number format/precision=4,
                xlabel=$\vert T_{pred}-T\vert$,
            }
        ]
        
            \edef\imagepath{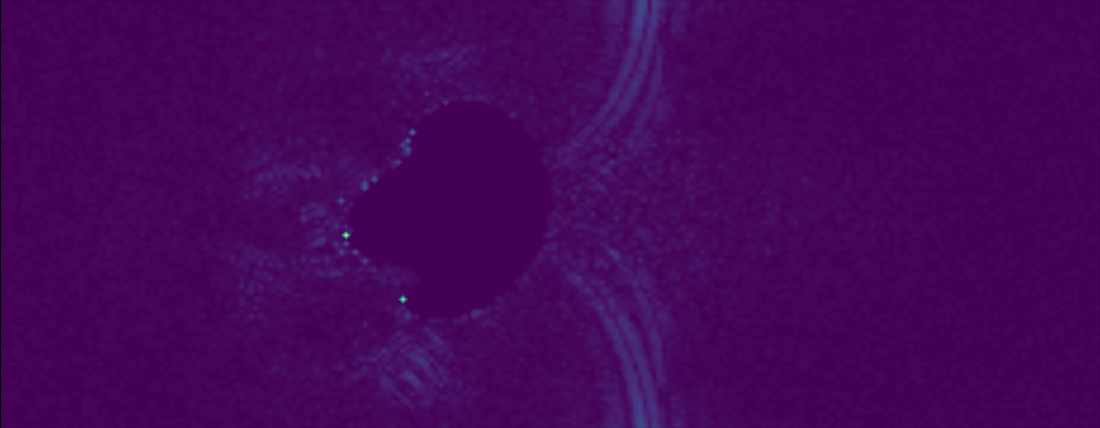}
            \addplot graphics[xmin=0e-4,xmax=7e-4,ymin=0e-4,ymax=3e-4]{\imagepath};
    
    \end{groupplot}
    \node[below = 1.5cm of my plots c1r1.south west,
        anchor=west,
    ] {(a) $T_{pred}$};
    \node[below = 1.5cm of my plots c2r1.south west,
        anchor=west,
    ] {(b) Point-wise error};
\end{tikzpicture}
    

%% file: sections_scaling.tex
\subsection{Modified scaling effect of temperature field}
\begin{figure}[tbph]
    \input{figures_ae_scaling.tex}
    \caption{Temporal (a) average and (b) standard deviation of HE temperature ($K$) for 36-case training data.}
    \label{fig:ae-scaling}
\end{figure}
Figure~\ref{fig:ae-scaling} shows the average and standard deviation in (\ref{eq:T-scale1})
for a training dataset of $N_p=36$ parameter points.
The parameter points are uniformly spaced as a $6\times6$ array in the range of $[\deg{0}, \deg{20}]\times[1\mu m, 1.2\mu m]$.
The initial pore shape distinctly appears in both the average and the standard deviation.
High variance is observed downstream of the initial pore location,
where the hot spot is being developed and ejected out of the pore collapse process.
We note that the average within the pore region exhibits high temperature
due to hot spot development following the pore collapse,
while the variance within the initial pore region is significantly low.
\refC{These} statistics cannot be captured well with the standard scaling in (\ref{eq:T-scale0}):
as all the $0K$ values are included, the temperature variance appears to be high in the pore region,
though it is not in reality.

%% file: figures_ae_scaling.tex
\begin{tikzpicture}[
    font=\small,
    spy using outlines={circle,black,magnification=6,size=1.5cm, connect spies}
    ]
    
    \pgfplotsset{set layers=standard}
    \begin{groupplot}[
        group style={
            group name = my plots,
            group size= 2 by 1,
            xlabels at =edge bottom,
            horizontal sep=3.5cm,
            vertical sep=2.cm,
        },
        name=chung,
    ]    

        \nextgroupplot[
            height = 0.22\textwidth,
            width = 0.45\textwidth,
            enlarge x limits={false, abs value = 0mm},
            ylabel={$x_2$ ($\mu m$)},
            xlabel={$x_1$ ($\mu m$)},
            tick scale binop ={\times},
            xmin=0,xmax=5,ymin=0,ymax=2,
            point meta min=436., point meta max=762,
            colorbar, colormap/viridis,
            colorbar style={
                font=\scriptsize,
                xticklabel pos=upper,
                scaled y ticks=false,
                ytick={436, 500, 600, 700, 762},
                yticklabels={436, 500, 600, 700, 762},
                /pgf/number format/precision=4,
                xlabel=$\overline{T}$,
            }
        ]
        
            \edef\imagepath{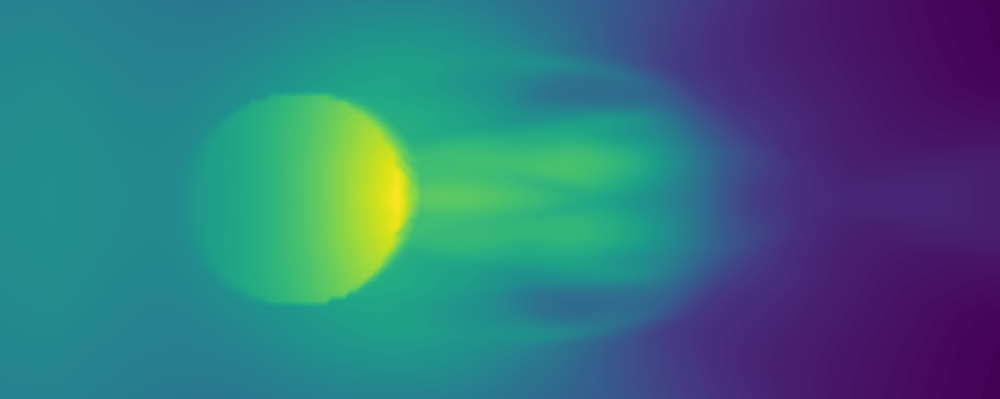}
            \addplot graphics[xmin=0,xmax=5,ymin=0,ymax=2]{\imagepath};


        \nextgroupplot[
            height = 0.22\textwidth,
            width = 0.45\textwidth,
            enlarge x limits={false, abs value = 0mm},
            ylabel={$x_2$ ($\mu m$)},
            xlabel={$x_1$ ($\mu m$)},
            tick scale binop ={\times},
            xmin=0,xmax=5,ymin=0,ymax=2,
            point meta min=8., point meta max=241.1760,
            colorbar, colormap/viridis,
            colorbar style={
                font=\scriptsize,
                xticklabel pos=upper,
                scaled y ticks=false,
                ytick={8., 100, 200, 240},
                yticklabels={8, 100, 200, 240},
                /pgf/number format/precision=4,
                xlabel=$\mathrm{std}[T]$,
            }
        ]
        
            \edef\imagepath{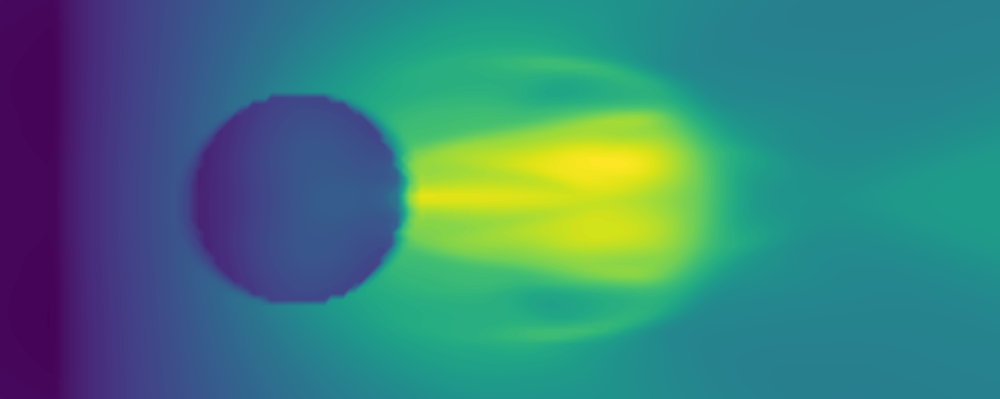}
            \addplot graphics[xmin=0,xmax=5,ymin=0,ymax=2]{\imagepath};
    
    \end{groupplot}
    \node[below = 1.5cm of my plots c1r1.south west,
        anchor=west,
    ] {(a) Average};
    \node[below = 1.5cm of my plots c2r1.south west,
        anchor=west,
    ] {(b) Standard deviation};
\end{tikzpicture}
    

%% file: figures_greedy_sampling.tex
\begin{tikzpicture}[
    font=\small,
    spy using outlines={circle,black,magnification=6,size=1.5cm, connect spies}
    ]
    
    \pgfplotsset{set layers=standard}
    \begin{groupplot}[
        group style={
            group name = my plots,
            group size= 2 by 2,
            xlabels at =edge bottom,
            horizontal sep=2.5cm,
            vertical sep=2.cm,
        },
        name=chung,
        height = 0.4\textwidth,
        width = 0.4\textwidth,
    ]    

        \nextgroupplot[
            ylabel={Major axis (micron)},
            xlabel={Angle (${}^\circ$)},
            tick scale binop ={\times},
            xmin=-1,xmax=21,ymin=0.99,ymax=1.21,
            xtick={0, 4, 8, 12, 16, 20},
            ytick={1, 1.04, 1.08, 1.12, 1.16, 1.2},
        ]
        
            \edef\imagepath{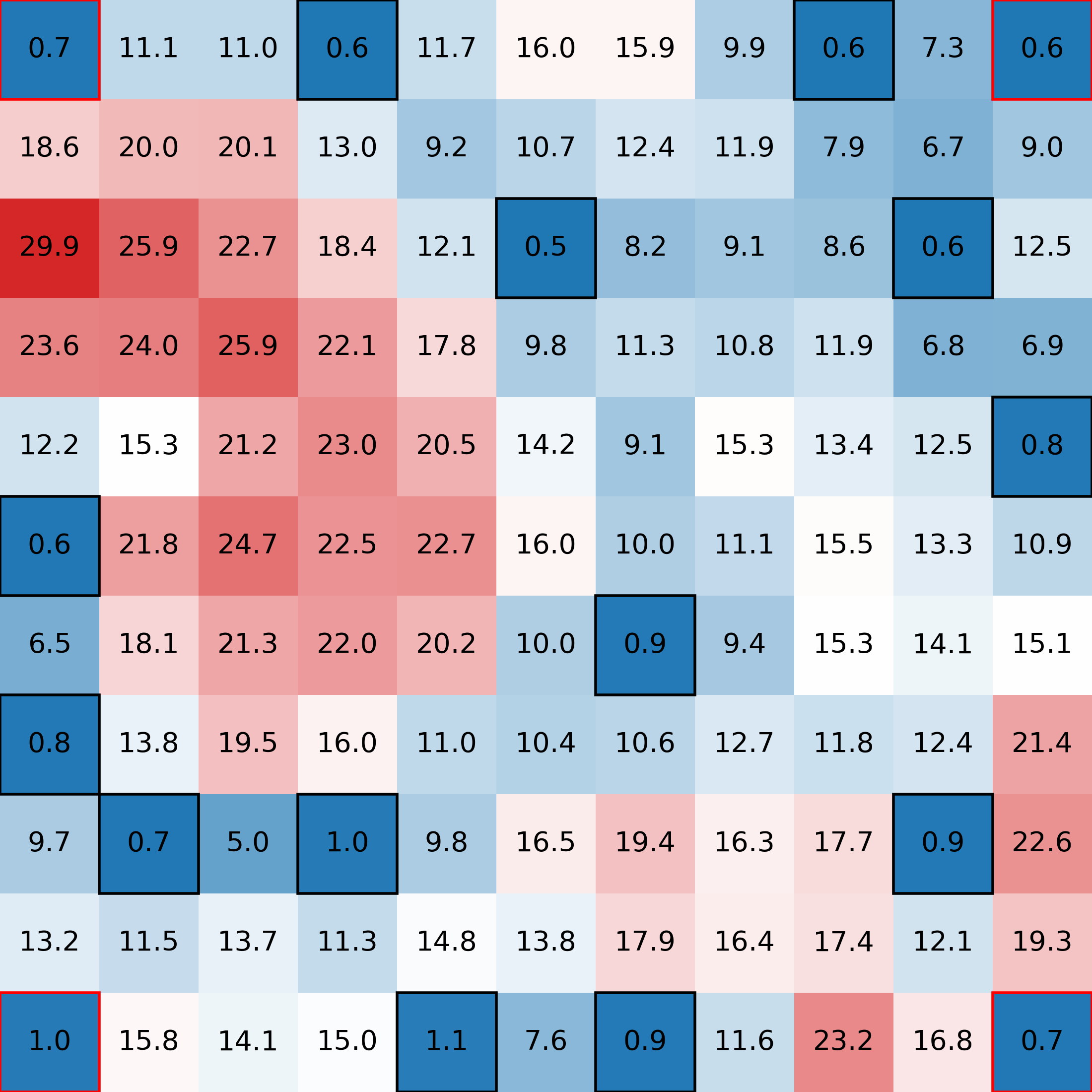}
            \addplot graphics[xmin=-1,xmax=21,ymin=0.99,ymax=1.21]{\imagepath};

        \nextgroupplot[
            ylabel={Major axis (micron)},
            xlabel={Angle (${}^\circ$)},
            tick scale binop ={\times},
            xmin=-1,xmax=21,ymin=0.99,ymax=1.21,
            xtick={0, 4, 8, 12, 16, 20},
            ytick={1, 1.04, 1.08, 1.12, 1.16, 1.2},
        ]
        
            \edef\imagepath{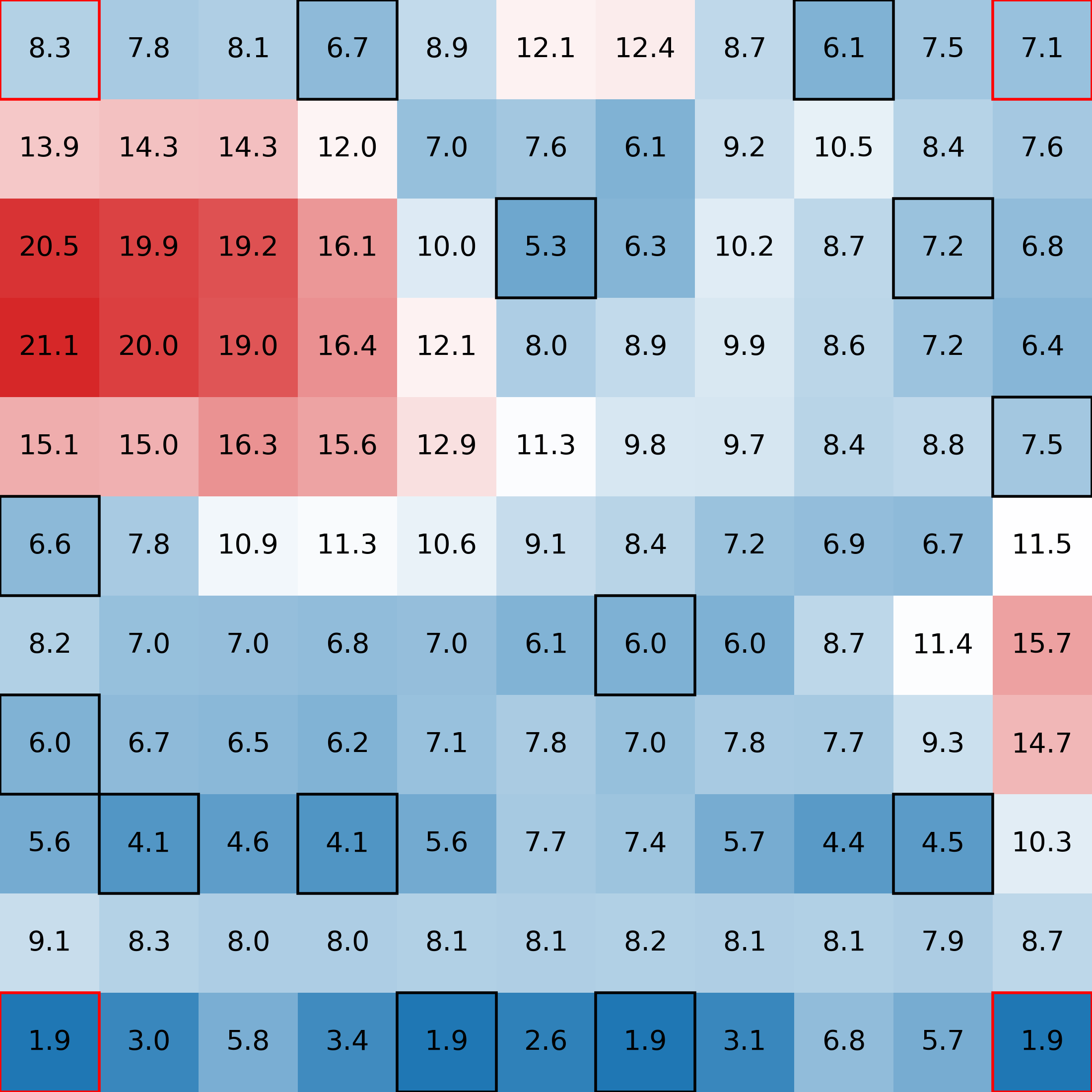}
            \addplot graphics[xmin=-1,xmax=21,ymin=0.99,ymax=1.21]{\imagepath};

        \nextgroupplot[
            ylabel={Major axis (micron)},
            xlabel={Angle (${}^\circ$)},
            tick scale binop ={\times},
            xmin=-1,xmax=21,ymin=0.99,ymax=1.21,
            xtick={0, 4, 8, 12, 16, 20},
            ytick={1, 1.04, 1.08, 1.12, 1.16, 1.2},
        ]
        
            \edef\imagepath{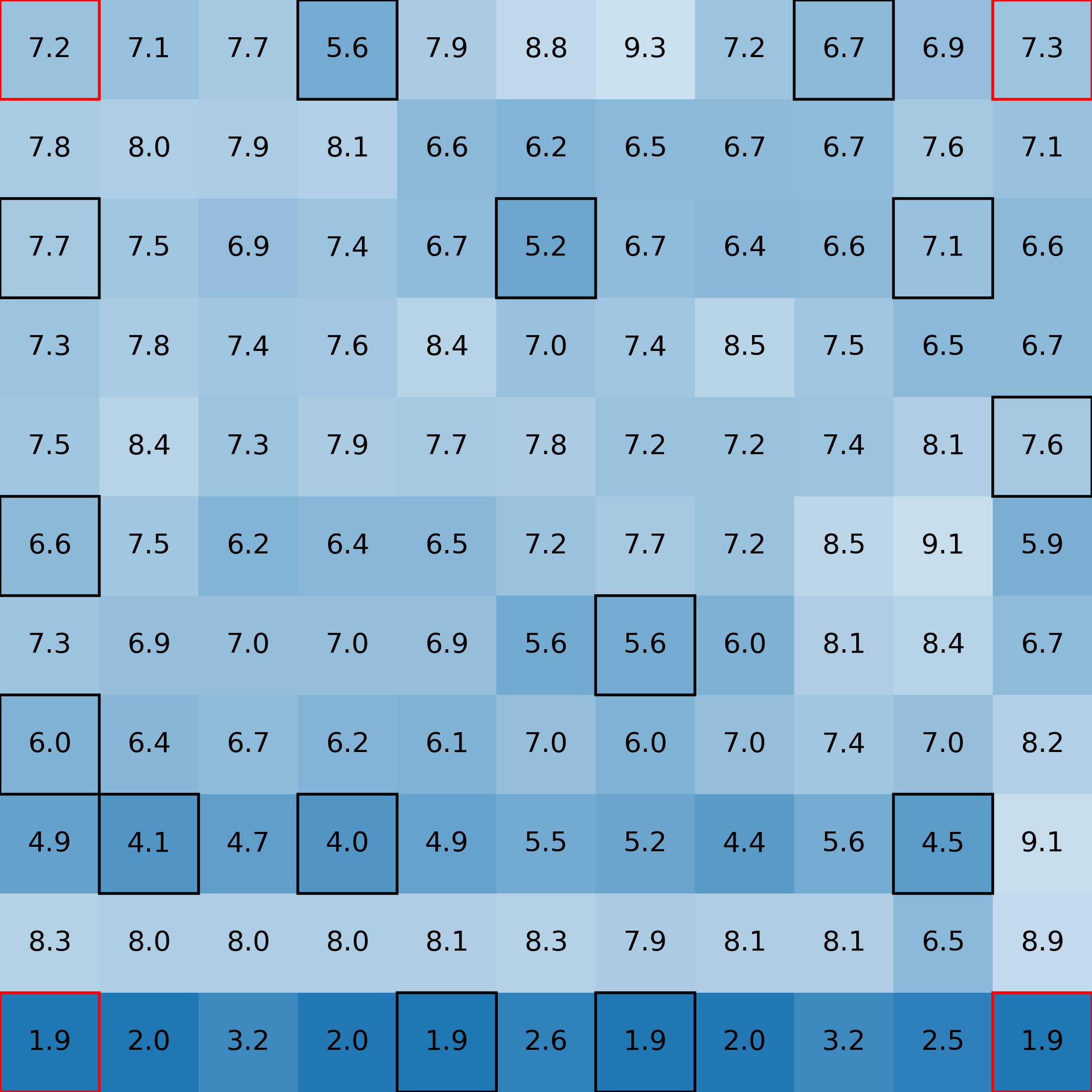}
            \addplot graphics[xmin=-1,xmax=21,ymin=0.99,ymax=1.21]{\imagepath};

        \nextgroupplot[
            ylabel={Major axis (micron)},
            xlabel={Angle (${}^\circ$)},
            tick scale binop ={\times},
            xmin=-1,xmax=21,ymin=0.99,ymax=1.21,
            xtick={0, 4, 8, 12, 16, 20},
            ytick={1, 1.04, 1.08, 1.12, 1.16, 1.2},
        ]
        
            \edef\imagepath{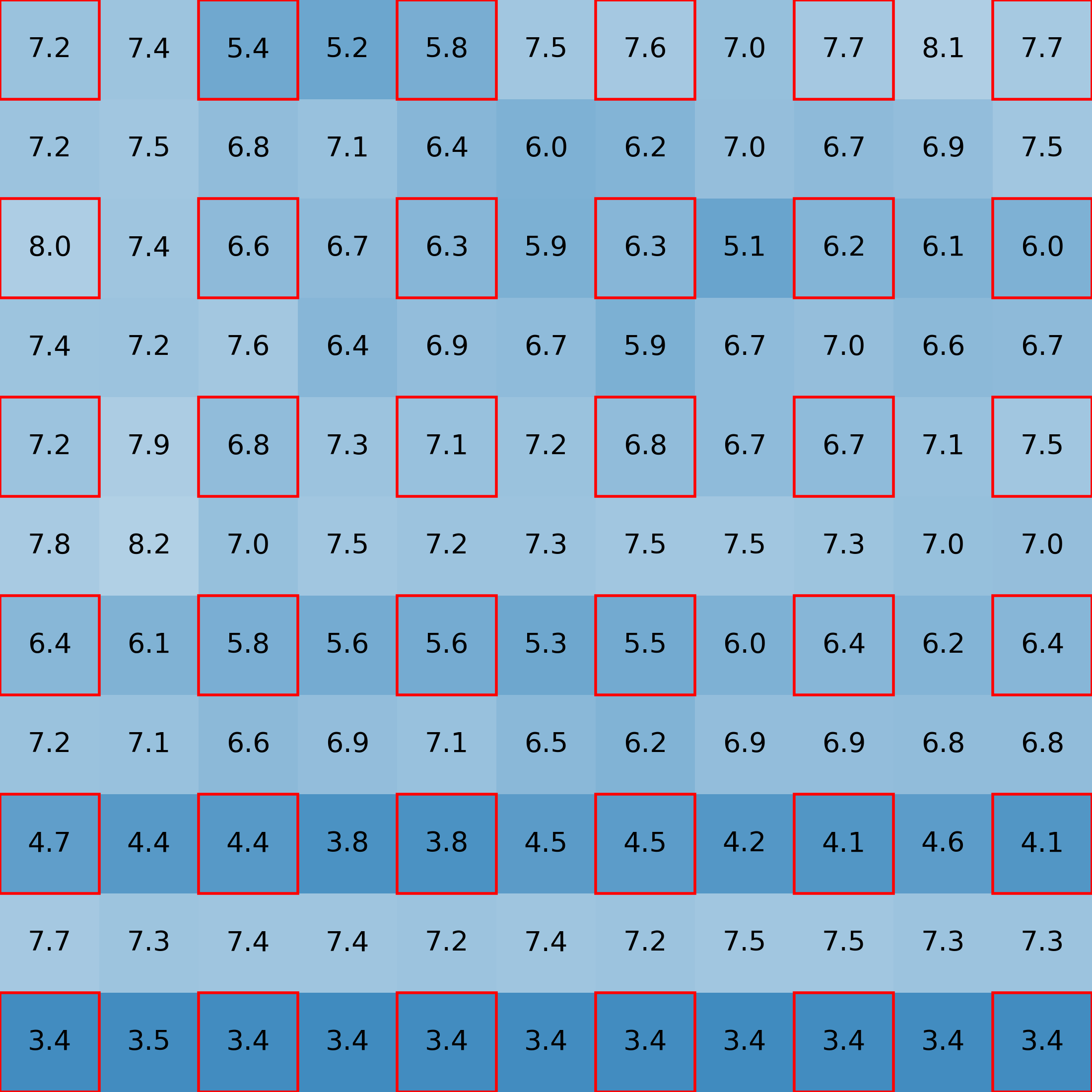}
            \addplot graphics[xmin=-1,xmax=21,ymin=0.99,ymax=1.21]{\imagepath};
    
    \end{groupplot}
    \node[below = 1.5cm of my plots c1r1.south west,
        anchor=west, xshift=-1cm,
    ] {(a) 14-th iteration, $\std_{gp}[\rT_{gp}](\bp)$ ($K$)};
    \node[below = 1.5cm of my plots c2r1.south west,
        anchor=west, xshift=-1cm,
    ] {(b) 14-th iteration, $\epsilon[\bT(\bp)]$ (\%)};
    \node[below = 1.5cm of my plots c1r2.south west,
        anchor=west, xshift=-1cm,
    ] {(c) 15-th iteration, $\epsilon[\bT(\bp)]$ (\%)};
    \node[below = 1.5cm of my plots c2r2.south west,
        anchor=west, xshift=-1cm,
    ] {(d) Uniform sampling, $\epsilon[\bT(\bp)]$ (\%)};
\end{tikzpicture}
    

%% file: figures_metrics.tex
\begin{tikzpicture}[
    font=\small,
    spy using outlines={circle,black,magnification=6,size=1.5cm, connect spies}
    ]
    
    \pgfplotsset{set layers=standard}
    \begin{groupplot}[
        group style={
            group name = my plots,
            group size= 3 by 1,
            xlabels at =edge bottom,
            horizontal sep=2.cm,
            vertical sep=2.cm,
        },
        name=chung,
        height = 0.3\textwidth,
        width = 0.3\textwidth,
    ]    

        \nextgroupplot[
            ylabel={Hot spot area ($\mu m^2$)},
            xlabel={Time $t$ ($ns$)},
            tick scale binop ={\times},
            xmin=0,xmax=1.85,
            legend style={
                draw=none, fill=none,
                at={(rel axis cs: 0., 1.0)},
                anchor=south west,
                legend columns=2,
            },
        ]

            \addplot+ [
                line width=1,
                smooth, solid, black,
                mark=none,
            ]
            table [x expr=\thisrowno{0} * 1e3, y expr=\thisrowno{1} * 1e2]{data_gplasdi_2params_3.metrics_truth.case13.iter014.txt};
            \addplot+ [
                line width=1,
                smooth, solid, blue,
                mark=none,
                mark options={fill=white,},
            ]
            table [x expr=\thisrowno{0} * 1e3, y expr=\thisrowno{1} * 1e2]{data_gplasdi_2params_3.metrics_pred.case13.iter014.txt};

            \legend{Truth, Prediction};

        \nextgroupplot[
            ylabel={Pore area ($\mu m^2$)},
            xlabel={Time $t$ ($ns$)},
            tick scale binop ={\times},
            xmin=0,xmax=1.85,
        ]

            \addplot+ [
                line width=1,
                smooth, solid, black,
                mark=none,
            ]
            table [x expr=\thisrowno{0} * 1e3, y expr=\thisrowno{2} * 1e2]{data_gplasdi_2params_3.metrics_truth.case13.iter014.txt};
            \addplot+ [
                line width=1,
                smooth, solid, blue,
                mark=none,
                mark options={fill=white,},
            ]
            table [x expr=\thisrowno{0} * 1e3, y expr=\thisrowno{2} * 1e2]{data_gplasdi_2params_3.metrics_pred.case13.iter014.txt};

        \nextgroupplot[
            ylabel={$T_{max}$ ($K$)},
            xlabel={Time $t$ ($ns$)},
            tick scale binop ={\times},
            xmin=0,xmax=1.85,
        ]

            \addplot+ [
                line width=1,
                smooth, solid, black,
                mark=none,
            ]
            table [x expr=\thisrowno{0} * 1e3, y index=3]{data_gplasdi_2params_3.metrics_truth.case116.iter014.txt};
            \addplot+ [
                line width=1,
                smooth, solid, blue,
                mark=none,
                mark options={fill=white,},
            ]
            table [x expr=\thisrowno{0} * 1e3, y index=3]{data_gplasdi_2params_3.metrics_pred.case116.iter014.txt};
    
    \end{groupplot}
    \node[below = 1.5cm of my plots c1r1.south west,
        anchor=west,
    ] {(a)};
    \node[below = 1.5cm of my plots c2r1.south west,
        anchor=west,
    ] {(b)};
    \node[below = 1.5cm of my plots c3r1.south west,
        anchor=west,
    ] {(c)};
\end{tikzpicture}
    

%% file: figures_prediction.tex
\begin{tikzpicture}[
    font=\scriptsize,
    spy using outlines={circle,black,magnification=6,size=1.5cm, connect spies}
    ]
    
    \pgfplotsset{set layers=standard}
    \begin{groupplot}[
        group style={
            group name = my plots,
            group size= 3 by 7,
            xlabels at =edge bottom,
            horizontal sep=2.3cm,
            vertical sep=1.8cm,
        },
        name=chung,
        height = 0.17\textwidth,
        width = 0.3\textwidth,
        colormap/viridis,
        colorbar style={
            font=\scriptsize,
            width=5pt,
            xticklabel pos=upper,
            scaled y ticks=false,
            /pgf/number format/precision=4,
            xlabel=$T$,
            xshift=-5pt,,
        }
    ]    

\nextgroupplot[
    enlarge x limits={false, abs value = 0mm},
    ylabel={$x_2$ ($\mu m$)},
    xlabel={$x_1$ ($\mu m$)},
    tick scale binop ={\times},
    xmin=0,xmax=5,ymin=0,ymax=2,
    ytick distance=1,
    point meta min=323., point meta max=606,
    colorbar,
]

    \edef\imagepath{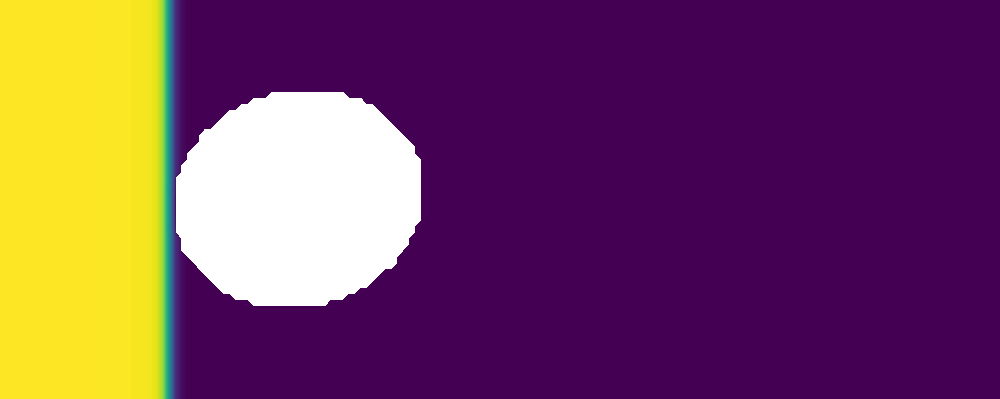}
    \addplot graphics[xmin=0,xmax=5,ymin=0,ymax=2]{\imagepath};

\nextgroupplot[
    enlarge x limits={false, abs value = 0mm},
    ylabel={$x_2$ ($\mu m$)},
    xlabel={$x_1$ ($\mu m$)},
    ytick distance=1,
    tick scale binop ={\times},
    xmin=0,xmax=5,ymin=0,ymax=2,
    point meta min=315, point meta max=623,
    colorbar,
]

    \edef\imagepath{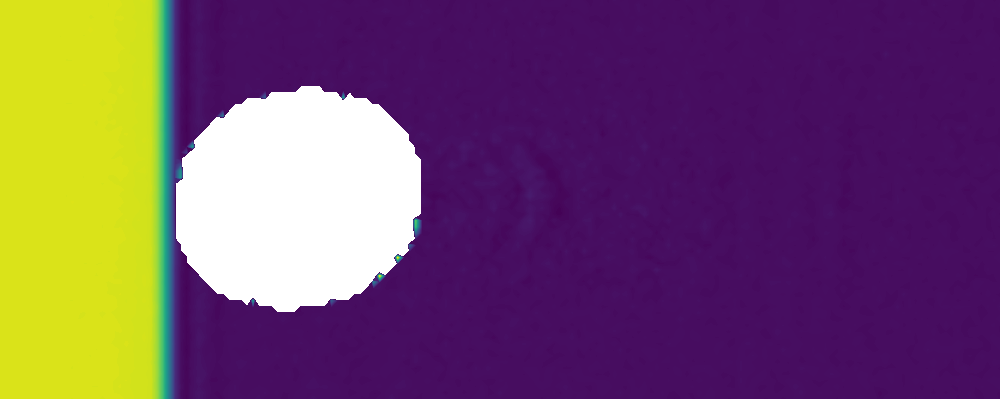}
    \addplot graphics[xmin=0,xmax=5,ymin=0,ymax=2]{\imagepath};

\nextgroupplot[
    enlarge x limits={false, abs value = 0mm},
    ylabel={$x_2$ ($\mu m$)},
    xlabel={$x_1$ ($\mu m$)},
    ytick distance=1,
    tick scale binop ={\times},
    xmin=0,xmax=5,ymin=0,ymax=2,
    point meta min=0., point meta max=83,
    colorbar,
    colorbar style={
        ytick distance=40,
    },
]

\edef\imagepath{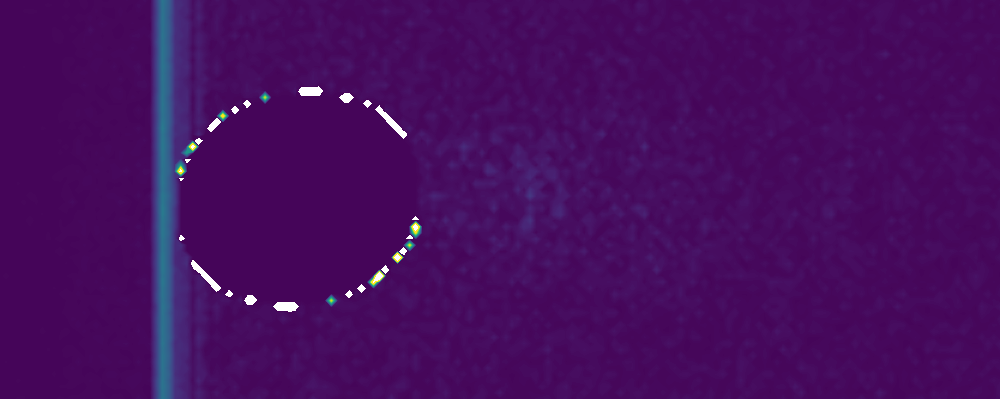}
\addplot graphics[xmin=0,xmax=5,ymin=0,ymax=2]{\imagepath};

\nextgroupplot[
    enlarge x limits={false, abs value = 0mm},
    ylabel={$x_2$ ($\mu m$)},
    xlabel={$x_1$ ($\mu m$)},
    ytick distance=1,
    tick scale binop ={\times},
    xmin=0,xmax=5,ymin=0,ymax=2,
    point meta min=323, point meta max=679,
    colorbar,
]

    \edef\imagepath{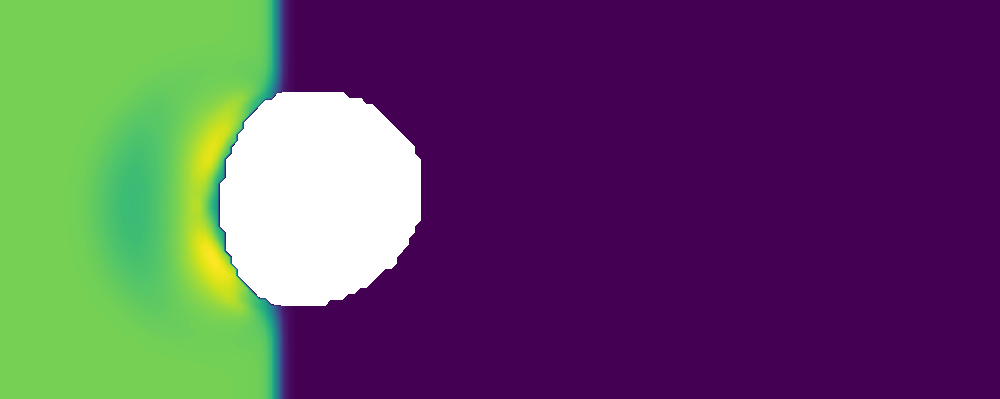}
    \addplot graphics[xmin=0,xmax=5,ymin=0,ymax=2]{\imagepath};

\nextgroupplot[
    enlarge x limits={false, abs value = 0mm},
    ylabel={$x_2$ ($\mu m$)},
    xlabel={$x_1$ ($\mu m$)},
    ytick distance=1,
    tick scale binop ={\times},
    xmin=0,xmax=5,ymin=0,ymax=2,
    point meta min=312., point meta max=673,
    colorbar,
]

    \edef\imagepath{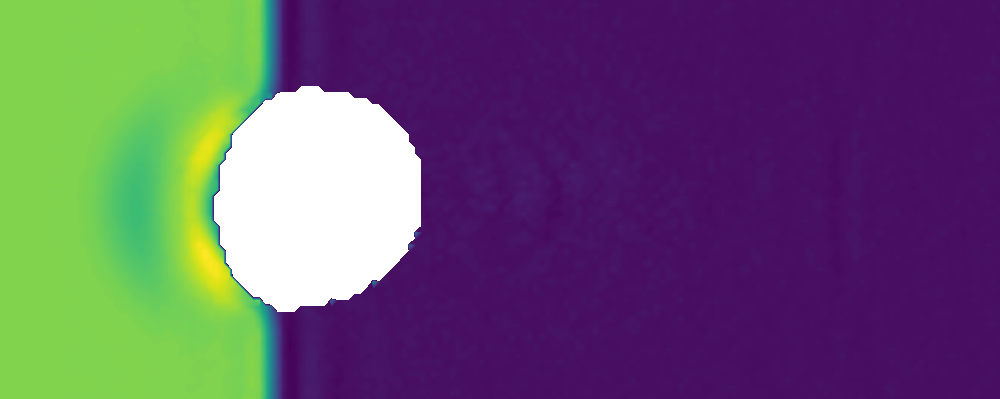}
    \addplot graphics[xmin=0,xmax=5,ymin=0,ymax=2]{\imagepath};

\nextgroupplot[
    enlarge x limits={false, abs value = 0mm},
    ylabel={$x_2$ ($\mu m$)},
    xlabel={$x_1$ ($\mu m$)},
    ytick distance=1,
    tick scale binop ={\times},
    xmin=0,xmax=5,ymin=0,ymax=2,
    point meta min=0., point meta max=96,
    colorbar,
    colorbar style={
        ytick distance=40,
    },
]

\edef\imagepath{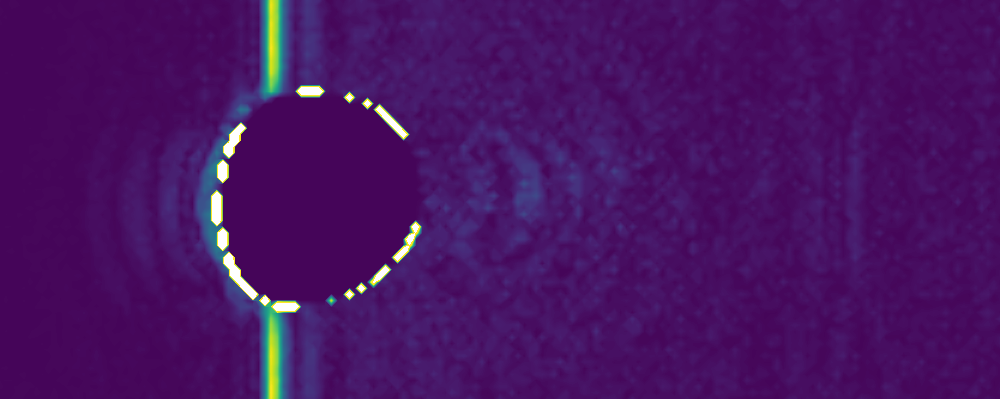}
\addplot graphics[xmin=0,xmax=5,ymin=0,ymax=2]{\imagepath};

\nextgroupplot[
    enlarge x limits={false, abs value = 0mm},
    ylabel={$x_2$ ($\mu m$)},
    xlabel={$x_1$ ($\mu m$)},
    ytick distance=1,
    tick scale binop ={\times},
    xmin=0,xmax=5,ymin=0,ymax=2,
    point meta min=323, point meta max=760,
    colorbar,
    colorbar style={
        ytick distance=150,
    },
]

    \edef\imagepath{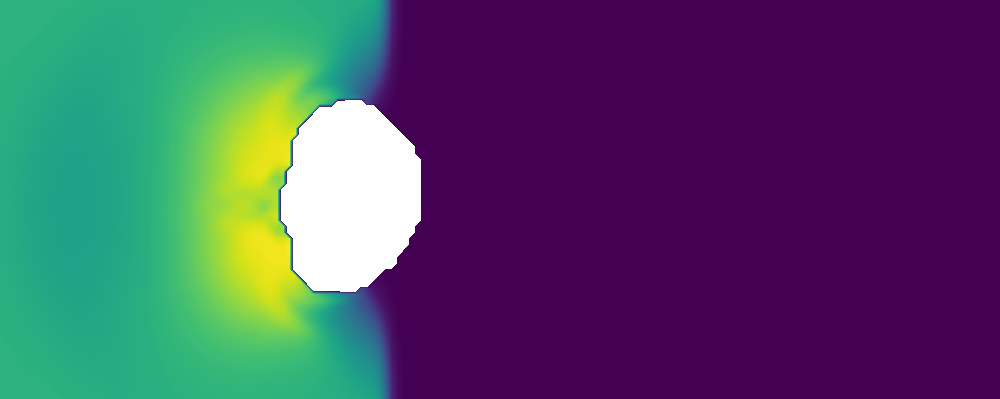}
    \addplot graphics[xmin=0,xmax=5,ymin=0,ymax=2]{\imagepath};

\nextgroupplot[
    enlarge x limits={false, abs value = 0mm},
    ylabel={$x_2$ ($\mu m$)},
    xlabel={$x_1$ ($\mu m$)},
    ytick distance=1,
    tick scale binop ={\times},
    xmin=0,xmax=5,ymin=0,ymax=2,
    point meta min=310., point meta max=752,
    colorbar,
    colorbar style={
        ytick distance=150,
    },
]

    \edef\imagepath{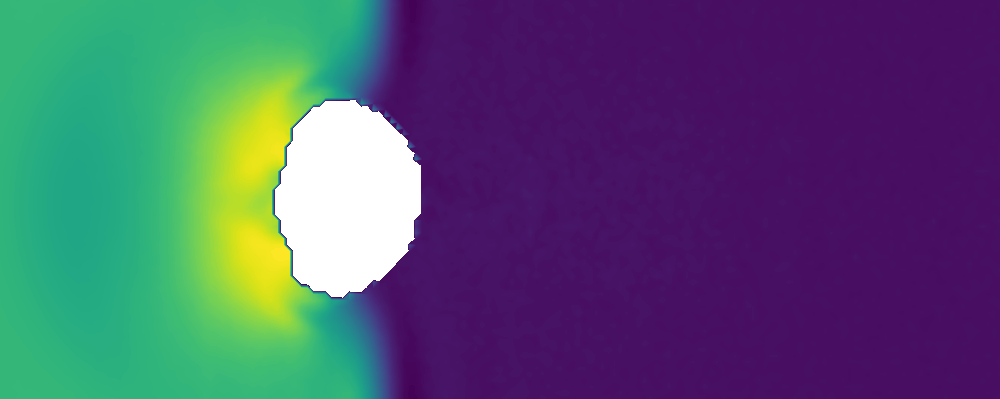}
    \addplot graphics[xmin=0,xmax=5,ymin=0,ymax=2]{\imagepath};

\nextgroupplot[
    enlarge x limits={false, abs value = 0mm},
    ylabel={$x_2$ ($\mu m$)},
    xlabel={$x_1$ ($\mu m$)},
    ytick distance=1,
    tick scale binop ={\times},
    xmin=0,xmax=5,ymin=0,ymax=2,
    point meta min=0., point meta max=90,
    colorbar,
    colorbar style={
        ytick distance=40,
    },
]

\edef\imagepath{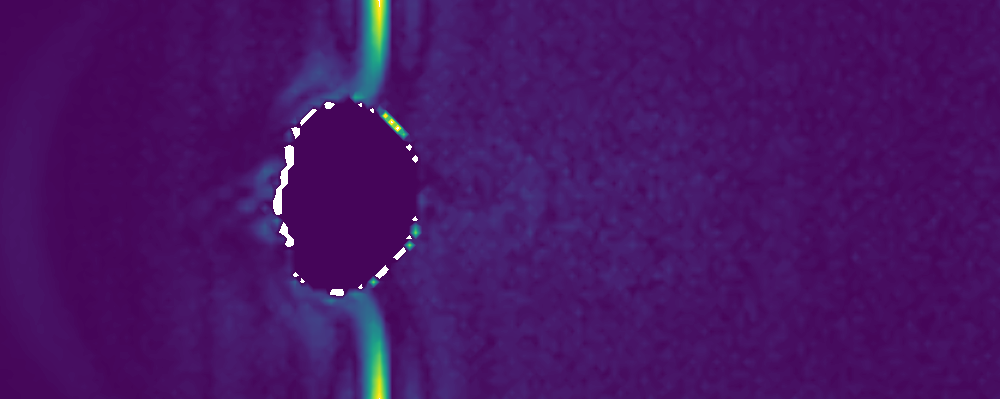}
\addplot graphics[xmin=0,xmax=5,ymin=0,ymax=2]{\imagepath};

\nextgroupplot[
    enlarge x limits={false, abs value = 0mm},
    ylabel={$x_2$ ($\mu m$)},
    xlabel={$x_1$ ($\mu m$)},
    ytick distance=1,
    tick scale binop ={\times},
    xmin=0,xmax=5,ymin=0,ymax=2,
    point meta min=323., point meta max=817,
    colorbar,
]

    \edef\imagepath{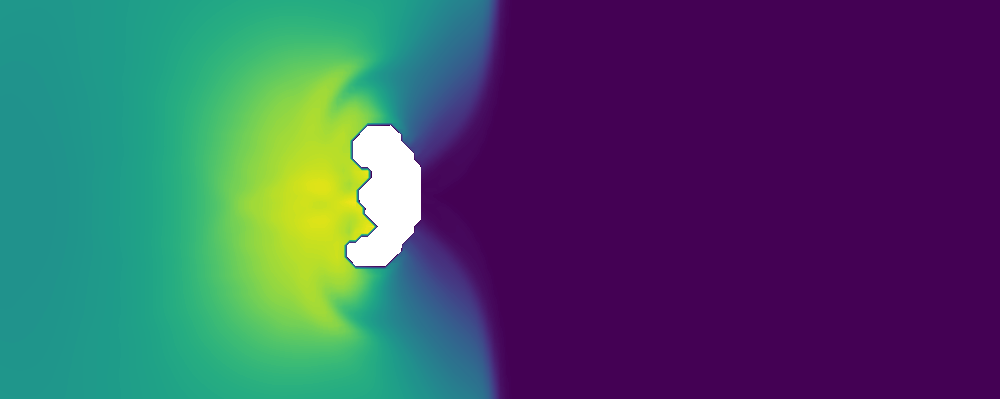}
    \addplot graphics[xmin=0,xmax=5,ymin=0,ymax=2]{\imagepath};

\nextgroupplot[
    enlarge x limits={false, abs value = 0mm},
    ylabel={$x_2$ ($\mu m$)},
    xlabel={$x_1$ ($\mu m$)},
    tick scale binop ={\times},
    ytick distance=1,
    xmin=0,xmax=5,ymin=0,ymax=2,
    point meta min=317., point meta max=787,
    colorbar,
    colorbar style={
        ytick={400, 600, 787},
        yticklabels={400, 600, 787},
    }
]

    \edef\imagepath{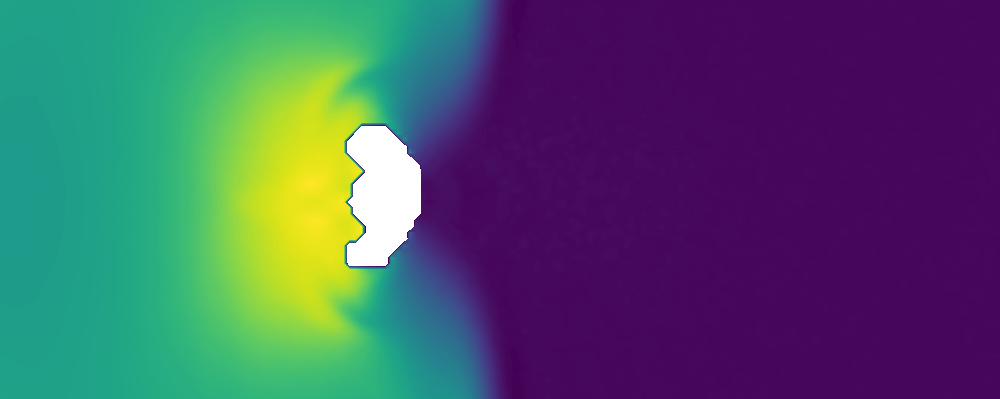}
    \addplot graphics[xmin=0,xmax=5,ymin=0,ymax=2]{\imagepath};

\nextgroupplot[
    enlarge x limits={false, abs value = 0mm},
    ylabel={$x_2$ ($\mu m$)},
    xlabel={$x_1$ ($\mu m$)},
    ytick distance=1,
    tick scale binop ={\times},
    xmin=0,xmax=5,ymin=0,ymax=2,
    point meta min=0., point meta max=72,
    colorbar,
    colorbar style={
        ytick distance=30,
    },
]

\edef\imagepath{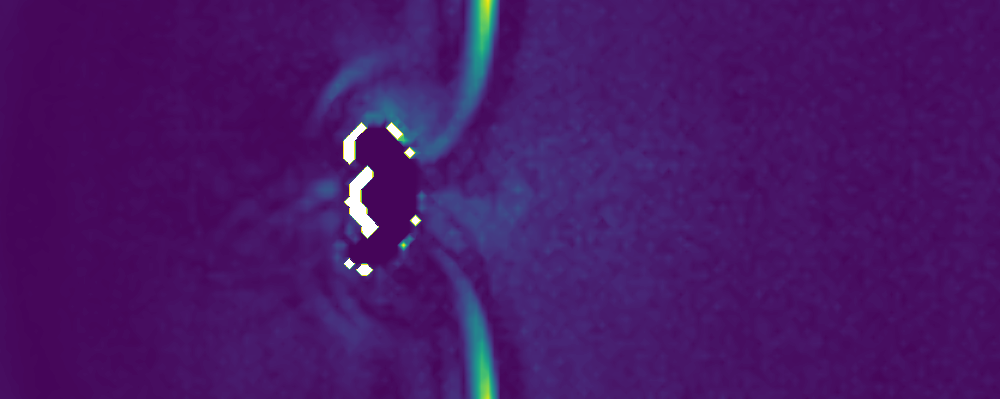}
\addplot graphics[xmin=0,xmax=5,ymin=0,ymax=2]{\imagepath};

\nextgroupplot[
    enlarge x limits={false, abs value = 0mm},
    ylabel={$x_2$ ($\mu m$)},
    xlabel={$x_1$ ($\mu m$)},
    ytick distance=1,
    tick scale binop ={\times},
    xmin=0,xmax=5,ymin=0,ymax=2,
    point meta min=323., point meta max=1337,
    colorbar,
    colorbar style={
        ytick distance=400,
    },
]

    \edef\imagepath{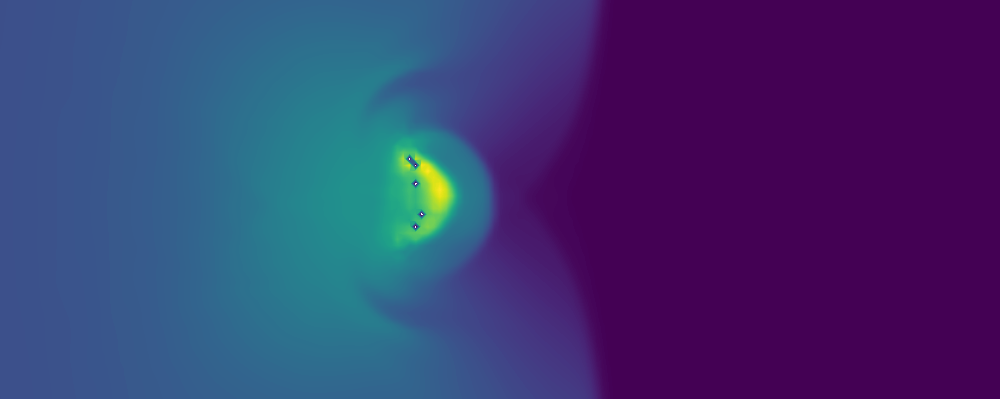}
    \addplot graphics[xmin=0,xmax=5,ymin=0,ymax=2]{\imagepath};

\nextgroupplot[
    enlarge x limits={false, abs value = 0mm},
    ylabel={$x_2$ ($\mu m$)},
    xlabel={$x_1$ ($\mu m$)},
    ytick distance=1,
    tick scale binop ={\times},
    xmin=0,xmax=5,ymin=0,ymax=2,
    point meta min=320., point meta max=1362,
    colorbar,
    colorbar style={
        ytick distance=400,
    },
]

    \edef\imagepath{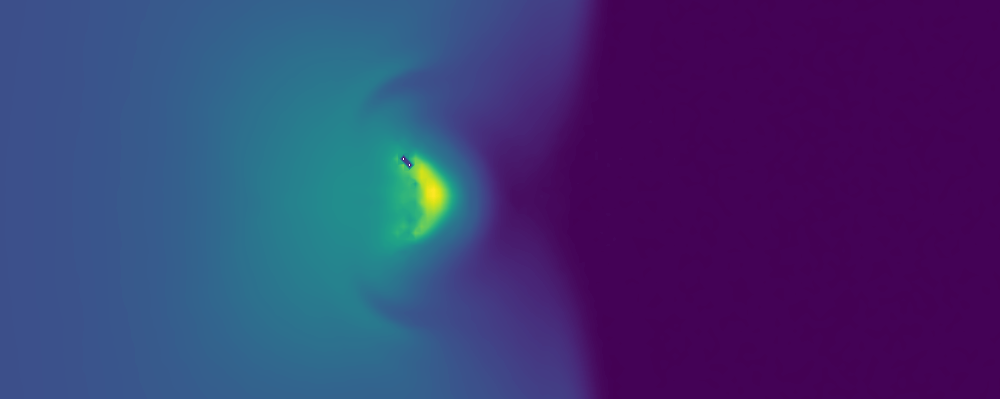}
    \addplot graphics[xmin=0,xmax=5,ymin=0,ymax=2]{\imagepath};

\nextgroupplot[
    enlarge x limits={false, abs value = 0mm},
    ylabel={$x_2$ ($\mu m$)},
    xlabel={$x_1$ ($\mu m$)},
    ytick distance=1,
    tick scale binop ={\times},
    xmin=0,xmax=5,ymin=0,ymax=2,
    point meta min=0., point meta max=193,
    colorbar,
    colorbar style={
        ytick distance=75,
    },
]

\edef\imagepath{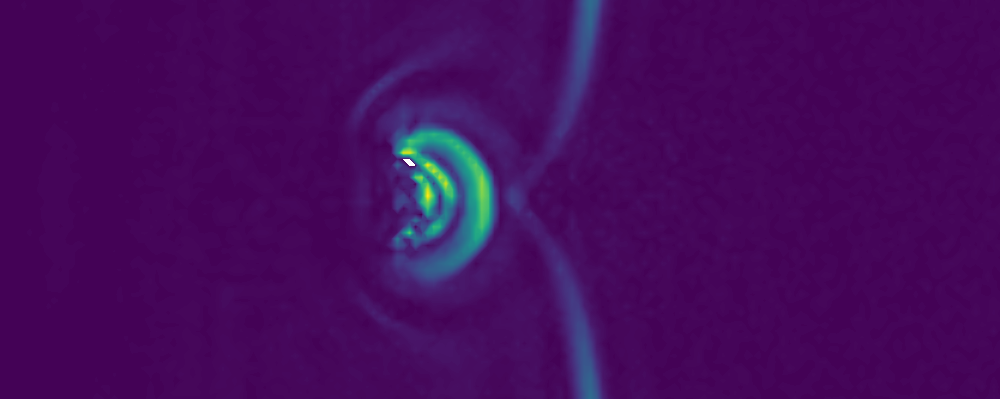}
\addplot graphics[xmin=0,xmax=5,ymin=0,ymax=2]{\imagepath};

\nextgroupplot[
    enlarge x limits={false, abs value = 0mm},
    ylabel={$x_2$ ($\mu m$)},
    xlabel={$x_1$ ($\mu m$)},
    ytick distance=1,
    tick scale binop ={\times},
    xmin=0,xmax=5,ymin=0,ymax=2,
    point meta min=323., point meta max=1179,
    colorbar,
    colorbar style={
        ytick distance=300,
    },
]

    \edef\imagepath{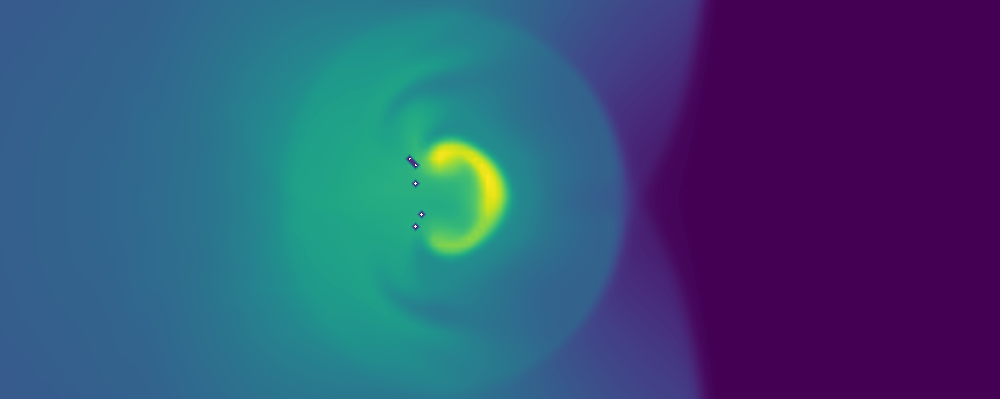}
    \addplot graphics[xmin=0,xmax=5,ymin=0,ymax=2]{\imagepath};

\nextgroupplot[
    enlarge x limits={false, abs value = 0mm},
    ylabel={$x_2$ ($\mu m$)},
    xlabel={$x_1$ ($\mu m$)},
    ytick distance=1,
    tick scale binop ={\times},
    xmin=0,xmax=5,ymin=0,ymax=2,
    point meta min=318., point meta max=1172,
    colorbar,
    colorbar style={
        ytick distance=300,
    },
]

    \edef\imagepath{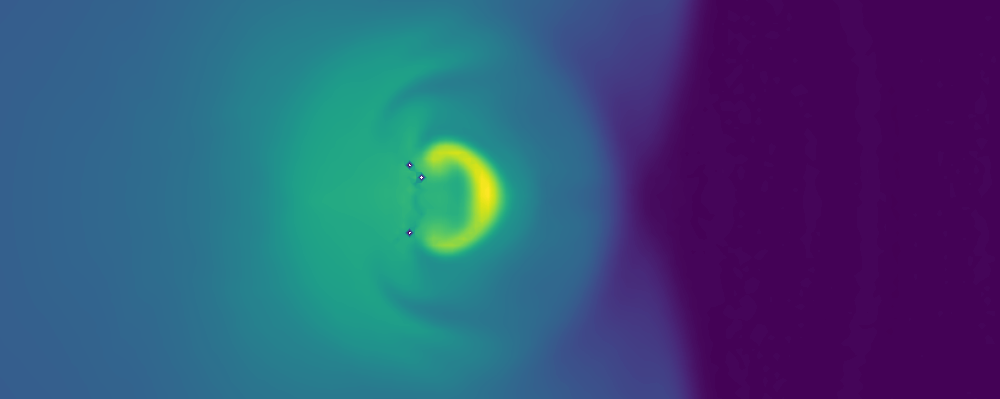}
    \addplot graphics[xmin=0,xmax=5,ymin=0,ymax=2]{\imagepath};

\nextgroupplot[
    enlarge x limits={false, abs value = 0mm},
    ylabel={$x_2$ ($\mu m$)},
    xlabel={$x_1$ ($\mu m$)},
    ytick distance=1,
    tick scale binop ={\times},
    xmin=0,xmax=5,ymin=0,ymax=2,
    point meta min=0., point meta max=147,
    colorbar,
]

\edef\imagepath{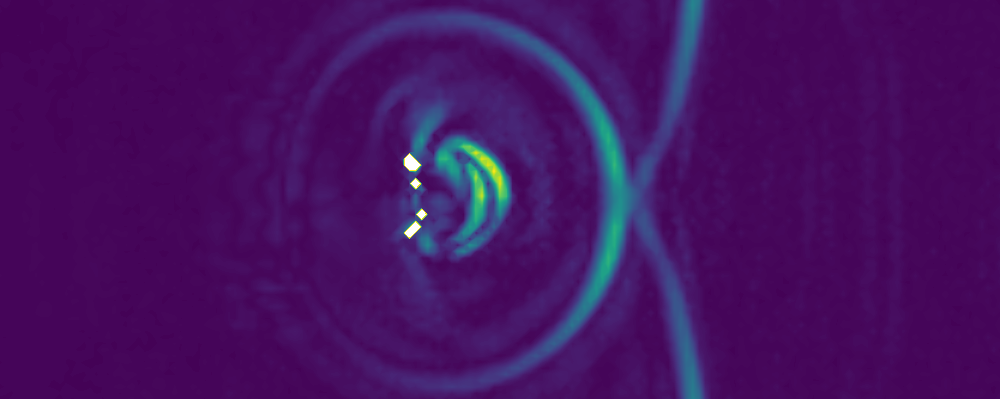}
\addplot graphics[xmin=0,xmax=5,ymin=0,ymax=2]{\imagepath};

\nextgroupplot[
    enlarge x limits={false, abs value = 0mm},
    ylabel={$x_2$ ($\mu m$)},
    xlabel={$x_1$ ($\mu m$)},
    ytick distance=1,
    tick scale binop ={\times},
    xmin=0,xmax=5,ymin=0,ymax=2,
    point meta min=542., point meta max=1027,
    colorbar,
]

    \edef\imagepath{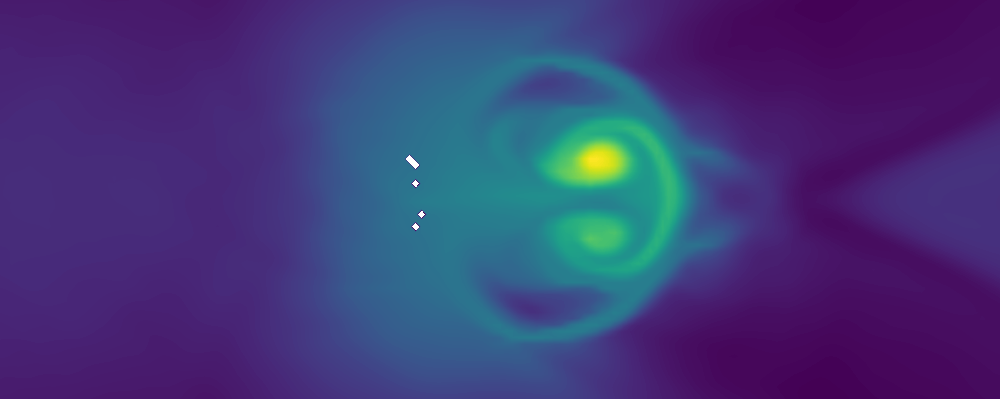}
    \addplot graphics[xmin=0,xmax=5,ymin=0,ymax=2]{\imagepath};

\nextgroupplot[
    enlarge x limits={false, abs value = 0mm},
    ylabel={$x_2$ ($\mu m$)},
    xlabel={$x_1$ ($\mu m$)},
    ytick distance=1,
    tick scale binop ={\times},
    xmin=0,xmax=5,ymin=0,ymax=2,
    point meta min=542., point meta max=1034,
    colorbar,
]

    \edef\imagepath{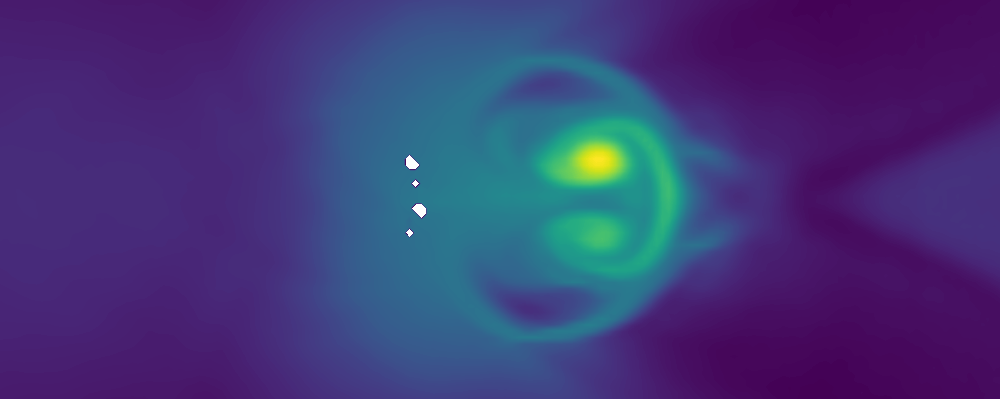}
    \addplot graphics[xmin=0,xmax=5,ymin=0,ymax=2]{\imagepath};

\nextgroupplot[
    enlarge x limits={false, abs value = 0mm},
    ylabel={$x_2$ ($\mu m$)},
    xlabel={$x_1$ ($\mu m$)},
    ytick distance=1,
    tick scale binop ={\times},
    xmin=0,xmax=5,ymin=0,ymax=2,
    point meta min=0., point meta max=32,
    colorbar,
]

\edef\imagepath{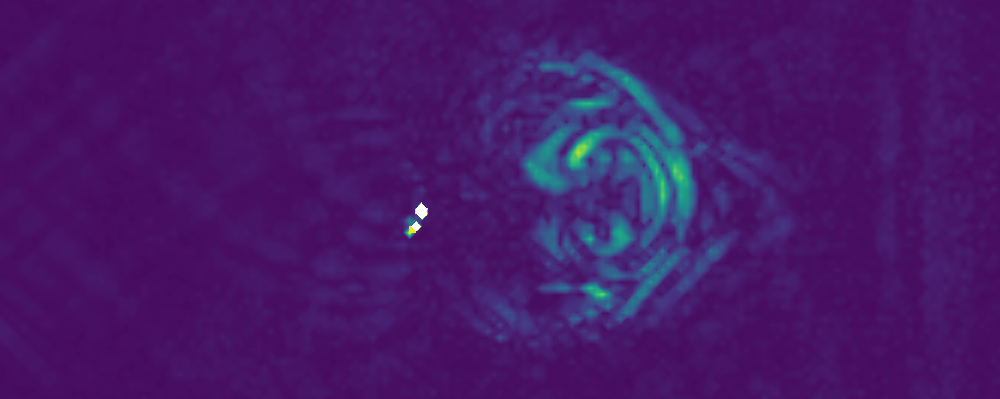}
\addplot graphics[xmin=0,xmax=5,ymin=0,ymax=2]{\imagepath};
    
    \end{groupplot}
    \node[below = 1.1cm of my plots c1r1.south west,
        anchor=west,
    ] {(a) $t=0.1 ns$};
    \node[below = 1.1cm of my plots c1r2.south west,
        anchor=west,
    ] {(b) $t=0.2 ns$};
    \node[below = 1.1cm of my plots c1r3.south west,
        anchor=west,
    ] {(c) $t=0.3 ns$};
    \node[below = 1.1cm of my plots c1r4.south west,
        anchor=west,
    ] {(d) $t=0.4 ns$};
    \node[below = 1.1cm of my plots c1r5.south west,
        anchor=west,
    ] {(e) $t=0.5 ns$};
    \node[below = 1.1cm of my plots c1r6.south west,
        anchor=west,
    ] {(f) $t=0.6 ns$};
    \node[below = 1.1cm of my plots c1r7.south west,
        anchor=west,
    ] {(g) $t=1.2 ns$};
\end{tikzpicture}
    

%% file: sections_conclusion.tex
\section{Conclusion}\label{sec:conclusion}

Modeling the dynamics of sharp, evolving interfaces, such as those encountered
in pore collapse phenomena, remains a significant challenge in reduced-order
modeling, particularly in data-scarce regimes. Traditional neural network-based
autoencoders struggle to accurately reconstruct such discontinuities due to
spectral bias and the inherent difficulty of embedding sharp interfaces into
smooth latent spaces.

In this work, we introduced \textit{LaSDI-IT}, a physics-aware extension of the
GPLaSDI framework designed for systems \kctwo{with} sharp interface
evolution. The core contribution lies in a modified autoencoder architecture
that simultaneously learns both the physical field (e.g., temperature) and a
binary indicator function that distinguishes material from pore regions. This
dual-target learning strategy mitigates the difficulty of regressing
discontinuous values and removes ambiguity in pore boundary reconstruction.

We applied LaSDI-IT to the problem of shock-induced pore collapse in high
explosives (HEs), where localized heating and interface deformation critically
influence hot spot formation. \kctwo{The latent dynamics are trained as
an ODE system via sparse regression and
are interpolated over parameter space using Gaussian process.}
The use of uncertainty-driven greedy sampling reduced the
required training data, offering predictive performance comparable to uniformly
sampled models at a significantly lower cost.

Our results demonstrate that LaSDI-IT provides accurate predictions of
essential physical metrics such as maximum temperature, pore area, and hot spot
extent. While peak temperatures are slightly underestimated due to smoothing
effects in latent space, the model captures the correct evolution and timing of
hot spot formation—key to understanding and designing HE material behavior.

Overall, this study demonstrates the potential of latent space dynamics
modeling to accelerate simulation-driven design for complex physical systems.
The LaSDI-IT framework is modular, data-efficient, and extensible to a variety
of interface-dominated phenomena, including multiphase flows, fracture
dynamics, and phase change processes. 

Future work will focus on extending the method to accommodate more complex pore
geometries, multi-pore interactions, and additional physical variables, such as
reactive chemistry and multi-material interfaces, to broaden its applicability
and robustness.

\par